\documentclass[useAMS,usenatbib]{mn2e}
\usepackage{graphicx}

\title[The electron temperatures]
      {The electron temperatures of SDSS high-metallicity 
giant extragalactic HII regions}
      
\author[L.S.Pilyugin, J.M.V\'{\i}lchez, B.Cedr\'es and T.X.Thuan]
       {L.S.~Pilyugin$^{1}$,  
        J.M.~V\'{\i}lchez$^{2}$,
        B.~Cedr\'es$^{2}$ 
        and T.X.~Thuan$^{3}$ \\
     $^{1}$ Main Astronomical Observatory
            of National Academy of Sciences of Ukraine,
            27 Zabolotnogo str., 03680 Kiev, Ukraine \\
     $^{2}$ Instituto de Astrof\'{\i}sica de Andaluc\'{\i}a,
            CSIC, Apdo, 3004, 18080 Granada, Spain \\
      $^{3}$ Astronomy Department, University of Virginia, P.O.Box 400325, 
              Charlottesville, VA 22904-4325     \\
             }

\date{Accepted 2009 Month  00. Received 2009 December 2; 
in original form 2009 October 26}

\begin{document}

\maketitle

\begin{abstract}
Spectra of high-metallicity (12+log(O/H) $\ga$ 8.2) H\,{\sc ii} regions 
where oxygen auroral lines are measurable in both the 
O$^+$  and O$^{++}$ zones, 
have been extracted from 
the Data Release 6 of the Sloan Digital Sky Survey (SDSS). 
Our final sample 
consists of 181 SDSS spectra of H\,{\sc ii} regions in  
galaxies in the redshift range from $\sim$0.025 to $\sim$0.17.  
The $t_{\rm 2,O}$--$t_{\rm 3,O}$  diagram is examined. 
  In the SDSS H\,{\sc ii} regions, the electron temperature 
$t_{\rm 2,O}$ is found 
to have a large scatter at a given value of the electron temperature $t_{\rm 3,O}$. 
The majority of the SDSS H\,{\sc ii} regions lie below the 
$t_{\rm 2,O}$--$t_{\rm 3,O}$ relation derived for  
H\,{\sc ii} regions in nearby galaxies, i.e. the positions of the SDSS 
H\,{\sc ii} regions show a systematic shift towards lower 
$t_{\rm 2,O}$ temperatures or/and towards higher 
$t_{\rm 3,O}$ temperatures. 
The scatter and shift of the SDSS H\,{\sc ii} regions 
in the $t_{\rm 2,O}$--$t_{\rm 3,O}$ diagram can be understood if  
they are composite nebulae excited by two or more 
ionizing sources of different temperatures.
\end{abstract}

\begin{keywords}

galaxies: abundances -- ISM: abundances -- H\,{\sc ii} regions

\end{keywords}

\section{Introduction}

%=====================

The determination of chemical abundances in cosmic objects started 150 
years ago, when Gustav Kirchhoff and Robert Bunsen discovered the laws of 
spectroscopy and founded the theory of spectrum analysis. 
In particular, they discovered that each atom exhibits a specific pattern of 
spectral lines, and that, therefore, by observing an object's spectrum, 
one can deduce, among other things, its composition and physical conditions. 
Kirchhoff applied spectral analysis techniques to study the chemical 
composition of the Sun. Many astronomers and physicists have made subsequent 
contributions to the theory of spectrum analysis. However, at the present time, 
the equations linking the abundance of a chemical element to the measured line
fluxes are still not beyond dispute.

Accurate oxygen abundances in H\,{\sc ii} regions can be derived via the 
classic $T_{\rm e}$ method, often referred to as the direct method. 
First, the electron temperature $t_3$ 
within the [O\,{\sc iii}] zone and the electron temperature $t_2$ within the 
[O\,{\sc ii}] zone are determined. Then the abundance is derived using the 
equations linking the ionic abundances to the measured line intensities and 
electron temperature. The ratio of nebular to auroral oxygen line intensities 
$Q_{\rm 3,O}$ = 
[O\,{\sc iii}]$\lambda$4959+$\lambda$5007/[O\,{\sc iii}]$\lambda 4363$ 
is used for the $t_{\rm 3}$ determination. 
 The ratio of nebular to auroral oxygen line intensities $Q_{\rm 2,O}$ = 
[O\,{\sc ii}]$\lambda$3727+$\lambda$3729/[O\,{\sc ii}]$\lambda$7320+$\lambda$7330
(or the ratio of nebular to auroral nitrogen line intensities 
[N\,{\sc ii}]$\lambda 6548+\lambda 6584$/[N\,{\sc ii}]$\lambda 5755$) 
is used for the $t_{\rm 2}$ determination. 
It is common practice to derive one value of the electron temperature 
$t_{\rm 3}$ or $t_{\rm 2}$ from the measured indicator $Q_{\rm 3,O}$ 
or $Q_{\rm 2,N}$ and to use the $t_{\rm 2}$--$t_{\rm 3}$ relation 
for the determination of the other electron temperature. 
Thus, in many cases, the abundances derived by the direct method are 
dependent on the adopted  $t_{\rm 2}$--$t_{\rm 3}$  relation. 
Those direct abundances are at the base of some recent empirical calibrations  
\citep{lcal,hcal,vybor,pilyuginthuan05,pettinipagel04}.  
So the abundances derived through those empirical calibrations 
are also  dependent on the adopted  $t_{\rm 2}$--$t_{\rm 3}$  relation.

H\,{\sc ii} region models are widely used to establish the 
$t_{\rm 2}$--$t_{\rm 3}$ relation. 
It is commonly accepted that there is one-to-one correspondence between 
the $t_{\rm 2}$ and $t_{\rm 3}$ electron temperatures.  
Several versions of a one-dimensional 
$t_{\rm 2}$--$t_{\rm 3}$ relation have been proposed. 
A widely used version is that by \citet{campbelletal86} 
(see also \citet{garnett92}) based on the H\,{\sc ii} region models of 
\citet{stasinska82}. 
Other relations have been proposed by \citet{pageletal92,izotovetal97}  
(based on H\,{\sc ii} region model calculations by \citet{stasinska90}),  
and \citet{deharveng00} (based on H\,{\sc ii} region model calculations 
by \citet{stasinskaschaerer97}). 
\citet{oey00} have found that the relation of \citet{campbelletal86} 
is reasonable for $t_{\rm 3}$ $>$ 1.0 ($t_{\rm 3}$ is in 
 units of 10$^4$K), however at lower temperatures, 
the models are more consistent with an isothermal nebula. 
We have derived a  $t_{\rm 2}$--$t_{\rm 3}$  relation based 
on the idea that the $T_{\rm e}$ method equation for O$^{++}$/H$^+$ applied to 
the O$^{++}$ zone and that for O$^{+}$/H$^+$ applied to the O$^{+}$ zone 
must result in the same value of the oxygen abundance \citep{pilyuginetal06b}. 

There have been several attempts to establish the  $t_{\rm 2}$--$t_{\rm 3}$ 
relation using direct measurements of both $t_{\rm 2}$ and 
$t_{\rm 3}$ in H\,{\sc ii} regions. 
\citet{kennicuttetal03} have used the measured $Q_{\rm 3,O}$ and $Q_{\rm 2,O}$ 
values  to derive $t_{\rm 3,O}$ and $t_{\rm 2,O}$, respectively, for a number of 
H\,{\sc ii} regions in nearby galaxies. Comparing $t_{\rm 2,O}$ to 
$t_{\rm 3,O}$, they found the surprising result that the two temperatures 
are uncorrelated for most of the objects in their sample. They noted that, 
while the exact cause of the absence of correlation is not known, a possible 
explanation is  the contribution of recombination processes to the 
population of the level giving rise to the 
[O\,{\sc ii}]$\lambda$7320+$\lambda$7330 lines. 
\citet{izotovetal06} found a correlation between the $t_{\rm 2,O}$ and 
$t_{\rm 3,O}$ derived for a sample of low-metallicity SDSS  H\,{\sc ii} 
regions. Their  $t_{\rm 2,O}$--$t_{\rm 3,O}$  relation follows that  
predicted by photoionization models, but the scatter of the data points is 
large. \citet{izotovetal06} attributed the large scatter to substantial  
flux errors of the weak [O\,{\sc ii}]$\lambda 7320+\lambda 7330$ emission 
lines. 

Another interpretation of the scatter in the  $t_{\rm 2}$--$t_{\rm 3}$  
diagram has been proposed. \citet {hageleetal06,hageleetal08} 
and \citet{pilyugin07} 
have suggested that there is not a  one-to-one correspondance between 
$t_{\rm 2}$ and $t_{\rm 3}$, but that  the  $t_{\rm 2}$--$t_{\rm 3}$ 
relation is dependent on some parameter such as electron density 
or excitation parameter. On the other hand, \citet{pilyuginetal09} have found 
that there is a one-to-one correspondence between $t_{\rm 2}$ and 
$t_{\rm 3}$ for H\,{\sc ii} regions in nearby galaxies with weak nebular 
$R_3$ lines (log$R_3$ $\la$ 0.5, see the definition of $R_3$ below). 
However, H\,{\sc ii} regions with strong nebular $ R_3$ lines (log$R_3$ $\ga$ 
0.5) do not follow this relation. As a result, the one-to-one correspondence 
between   $t_{\rm 2}$ and $t_{\rm 3}$ disappears if a sample contains  
H\,{\sc ii} regions with both weak and strong $R_3$ lines.  

We wish to study here  the  $t_{\rm 2,O}$--$t_{\rm 3,O}$  relation, using 
a large sample of  H\,{\sc ii} regions extracted from the Sloan Digital 
Sky Survey (SDSS). The paper is organised as follows. The SDSS H\,{\sc ii} 
region sample  is described in Section 2. The  $t_{\rm 2,O}$--$t_{\rm 3,O}$  
diagram is discussed in Section 3.  The origin of the scatter in that diagram 
is examined in Section 4. Section 5 presents the conclusions.

Throughout the paper, we will be using the following notations for the line 
fluxes: \\ 
$R$ = [O\,{\sc iii}]$\lambda$4363 = $I_{{\rm [OIII]} \lambda 4363} /I_{{\rm H}\beta }$,  \\
$R_2$ = [O\,{\sc ii}]$\lambda$3727+$\lambda$3729  
      = $I_{[OII] \lambda 3727+ \lambda 3729} /I_{{\rm H}\beta }$,  \\
$R_3$ = [O\,{\sc iii}]$\lambda$4959+$\lambda$5007        = 
$I_{{\rm [OIII]} \lambda 4959+ \lambda 5007} /I_{{\rm H}\beta }$,  \\
\( \mbox{[O\,{\sc ii}]}\lambda 7325  
      = \mbox{[O\,{\sc ii}]} \lambda 7320+ \lambda 7330  
      =  I_{{\rm [OII]} \lambda 7320+ \lambda 7330} /I_{{\rm H}\beta } , \) \\ 
With these definitions,  the excitation parameter $P$ can be expressed as:   \\
$P$ = $R_3$/($R_2$+$R_3$), \\           
and the temperature indicators $Q_{2,O}$ and 
$Q_{\rm 3, O}$ can be expressed as:  \\ 
$Q_{\rm 2, O}$ = $R_2$/[O\,{\sc ii}]$\lambda$7325,  \\  
$Q_{\rm 3,O}$ = $R_3$/$R$. \\
The electron temperatures will be given in units of 10$^4$K.

\section{Sample description}

%=================

%***************************

\subsection{Galaxy selection}

%***************************

We construct our sample by selecting from the Data Release 6 of the SDSS  
spectra of H\,{\sc ii} regions which satisfy the following criteria: 
1) the two auroral lines  [O\,{\sc iii}]$\lambda$4363 
and [O\,{\sc ii}]$\lambda$7320+$\lambda$7330 are detected;  
2) the spectra have smooth line profiles. Particular attention was paid to 
the weak [O\,{\sc iii}]$\lambda$4363 and 
[O\,{\sc ii}]$\lambda$7320+$\lambda$7330 auroral lines  
since the accuracy of the electron 
temperature determination depends mainly  
on the measurement uncertainties of those emission lines; 
3) the emission lines do not have a broad component, thus excluding all AGNs.

%====================================    Fig  No 1            Kauffman 
\begin{figure}
\resizebox{1.00\hsize}{!}{\includegraphics[angle=000]{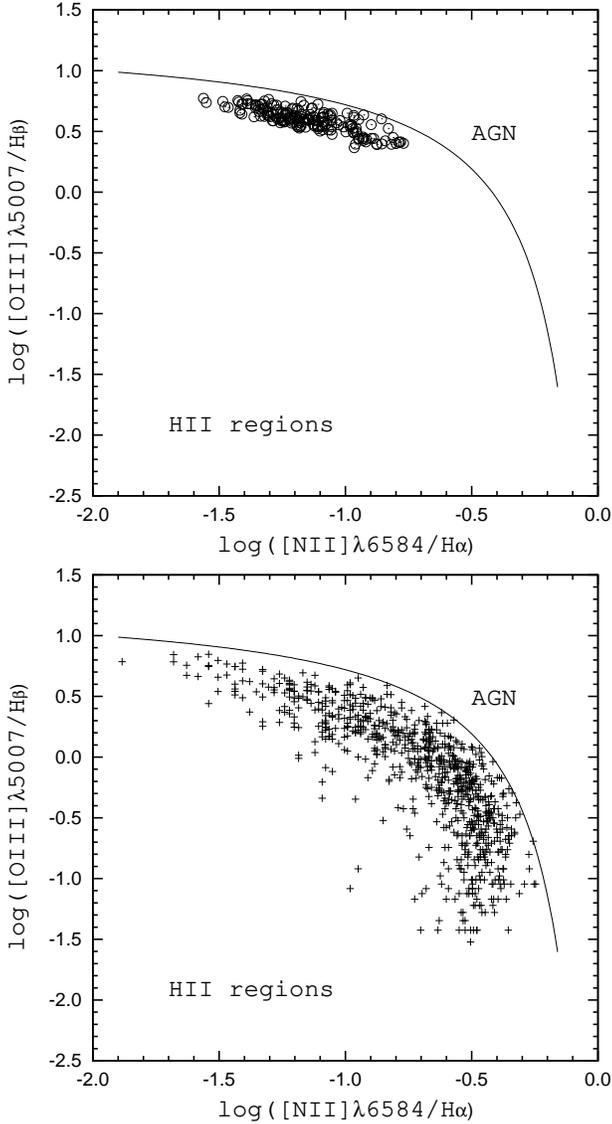}}
\caption{The N\,{\sc ii}]$\lambda$6584/H$\alpha$ versus [O\,{\sc iii}]$\lambda$5007/H$\beta$ diagram 
for our sample of  SDSS H\,{\sc ii} regions ({\it upper panel}) and 
for H\,{\sc ii} regions in nearby galaxies 
(a compilation of data from \citet{pilyuginetal04}) ({\it lower panel}).
The continuous line separates objects with a H\,{\sc ii} region-like 
spectrum from those containing an active galactic nucleus 
\citep{kauffmannetal03}.}
\label{figure:kauffmann}
\end{figure}

%====================================       Fig  No 2      Ne  [SII]6717/[SII]6731
\begin{figure}
\resizebox{1.00\hsize}{!}{\includegraphics[angle=000]{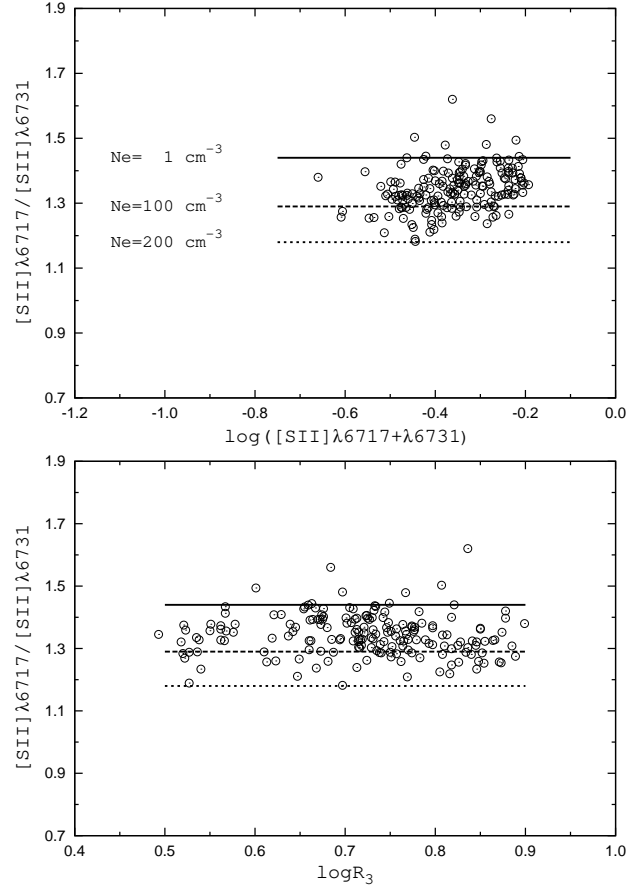}}
\caption{The density-sensitive [S\,{\sc ii}]$\lambda$6717/[S\,{\sc ii}]$\lambda$6731 
line ratio versus 
the [S\,{\sc ii}]$\lambda$6717+$\lambda$6731 emission line flux
({\it top panel}) and 
versus the $R_3$ emission line flux ({\it bottom panel}).
The open circles show the  SDSS H\,{\sc ii} regions. 
The solid line shows the zero-density limit ($N_{\rm e}$=1 cm$^{-3}$), 
the dashed line corresponds to the electron density $N_{\rm e}$=100 cm$^{-3}$, 
and the dotted line corresponds to the electron density $N_{\rm e}$=200 cm$^{-3}$.}
\label{figure:ne}
\end{figure}

Applying these criteria, we selected around 400 spectra 
out of the original 1 200 000 in the DR6. 
Although we visually inspected all  
spectra, we do not claim any completeness for our sample. 
The line intensities in the selected SDSS spectra of the H\,{\sc ii} regions have
been measured in the way described in \citet{pilyuginthuan07}.
In brief, it can be illustrated by the following example concerning the 
H${\alpha}$ line. The continuum flux level in the wavelength range from 
$\lambda_a$ = 6500\AA\ to  $\lambda_b$ = 6650\AA\  is approximated by the 
linear expression 
\begin{eqnarray}
f_{c}(\lambda)= c_0 + c_1 \lambda.
\label{equation:contin}
\end{eqnarray}
The values of the coefficients in Eq.(\ref{equation:contin}) are derived by 
an iteration procedure. In the first step, the H${\alpha}$ 
line region,  from 6540\AA\ to 6590\AA\ is excluded,  and all other data points 
are used to derive a first set of coefficients, using the least-squares method. Then, 
the point with the largest deviation is rejected, and a new set of coefficients 
is derived. The iteration procedure is continued until the differences 
between two successive values of $f_{\rm_c}$($\lambda_{\rm a}$) 
and $f_{\rm c}$($\lambda_{\rm b}$) are 
less than 0.01$f_{\rm c}$($\lambda_{\rm a}$) 
and 0.01$f_{\rm c}$($\lambda_{\rm b}$) respectively.

The profile of each line is approximated by a Gaussian of the form
\begin{eqnarray}
f(\lambda)= F\, \frac{1}{\sqrt{2\pi}\,\sigma} 
e^{-(\lambda - \lambda_0)^2/2\sigma^2}  ,
\label{equation:gauss}
\end{eqnarray}
where $\lambda_0$ is the central wavelength, $\sigma$ is the width of 
the line, and $F$ is the flux in the emission line. If there is  
absorption, then this line is fitted by two Gaussians simultaneously. In this case,  
the total flux at a fixed value of $\lambda$ is given by the expression
\begin{eqnarray}
       \begin{array}{lll}
f(\lambda) & = & f_{{\rm H}\alpha , {\rm em}}(\lambda) + 
f_{{\rm H}\alpha , {\rm abs}}(\lambda) + f_{\rm c}(\lambda) .
       \end{array}
\label{equation:ftot}   
\end{eqnarray}
The values of 
$F({\rm H}\alpha,{\rm em})$, $\lambda_0({\rm H}\alpha,{\rm em})$, 
$\sigma({\rm H}\alpha,{\rm em})$, 
$F({\rm H}{\alpha},{\rm abs})$, $\lambda_0({\rm H}{\alpha},{\rm abs})$, 
$\sigma({\rm H}{\alpha},{\rm abs})$, 
 are derived by requiring the mean difference  
\begin{eqnarray}
\epsilon = \sqrt{ \frac{1}{\rm n} \sum_{{\rm k}=1}^{{\rm k}={\rm n}} 
(f(\lambda_{\rm k})-f^{\rm obs}(\lambda_{\rm k}))^2}
\label{equation:df}
\end{eqnarray}
between the measured flux $f^{\rm obs}(\lambda_{\rm k})$ and the flux $f(\lambda_{\rm k}$) 
given by Eq.(\ref{equation:ftot}) to be minimum in the range $\lambda_{\rm a}$ --  
$\lambda_{\rm b}$. A similar procedure is adopted in the cases of  overlapping 
lines (e.g. [O\,{\sc ii}]$\lambda$3727 and [O\,{\sc ii}]$\lambda$3729). 

In the majority of cases, the profile of each line is well fitted by a Gaussian
since only spectra with smooth line profiles were selected. Thus, the 
uncertainties in the continuum level determination seem to make a 
dominant contribution to the uncertainties in the measurements of the 
weak auroral lines. 
The value of $\epsilon$ (Eq.(\ref{equation:df})) is 
the mean uncertainty in the measurements of the flux in a single 
spectral interval. If we interpret this value as 
the uncertainty in the continuum level determination, 
the formal uncertainty in the measurements of the weak line intensity 
due to the uncertainty in the continuum level determination 
can be estimated as $\epsilon$$_{\rm F}$ = $n$$\epsilon$, 
where $n$ is the number of points within the line profile.

Alternatively, the formal uncertainty in the weak line 
measurements defined in such a way can be interpreted as if the value of the 
continuum level is derived exactly and each point within the line profile 
involves an error equal to the mean uncertainty $\epsilon$ and all the 
errors have the same sign. 
One can expect that the formal uncertainty defined as $\epsilon$$_{\rm F}$ = 
$n$$\epsilon$ is close to the maximum possible error and that the 
real uncertainty 
in the line measurements is appreciably lower than this value. 
Consequently, the condition $\epsilon$$_{\rm F}$ $<$ 0.5, when applied to both 
auroral lines [O\,{\sc iii}]$\lambda$4363 and [O\,{\sc ii}]$\lambda$7325,
should select only 
objects with relatively precise line measurements, even though some 
objects with good measurements will be lost. 

The measured emission fluxes are then corrected for interstellar reddening 
using the theoretical H$\alpha$ to H$\beta$ ratio and the analytical 
approximation to the Whitford  interstellar reddening law from 
\citet{izotovetal94}. In several cases, the derived 
value of the extinction c(H$\beta$) is negative and is set to zero.

The dereddened line intensities of the H\,{\sc ii} regions in 
the final list are 
given in  Table~\ref{table:list} of the Appendix, available on line.
The line intensities are given on a scale in which I(H$\beta$) = 1.
The objects are listed in order of right ascension. 
The first column gives the order number of the object. 
The J2000.0 right ascension and declination of each object 
are given in columns 2 and 3 respectively. They are the SDSS identificators 
of each object. Right ascensions are in units of 
hours, minutes and seconds and declinations are in units 
of degrees, arcminutes, and arcseconds. 
The SDSS spectrum number, (composed of the 
plate number, the modified Julian date of observations and the  
number of the fiber on the plate) is listed in column 4.
The measured 
[O\,{\sc ii}]$\lambda$3727+$\lambda$3729,
[O\,{\sc iii}]$\lambda$4363, 
[O\,{\sc iii}]$\lambda$4959+$\lambda$5007, 
[S\,{\sc ii}]$\lambda$6717, 
[S\,{\sc ii}]$\lambda$6731, 
[O\,{\sc ii}]$\lambda$7320+$\lambda$7330 
line intensities are given in columns from 5 to 10 respectively.

%====================================    Fig  No  3    logR3  -  logR2    
\begin{figure*}
\resizebox{1.00\hsize}{!}{\includegraphics[angle=000]{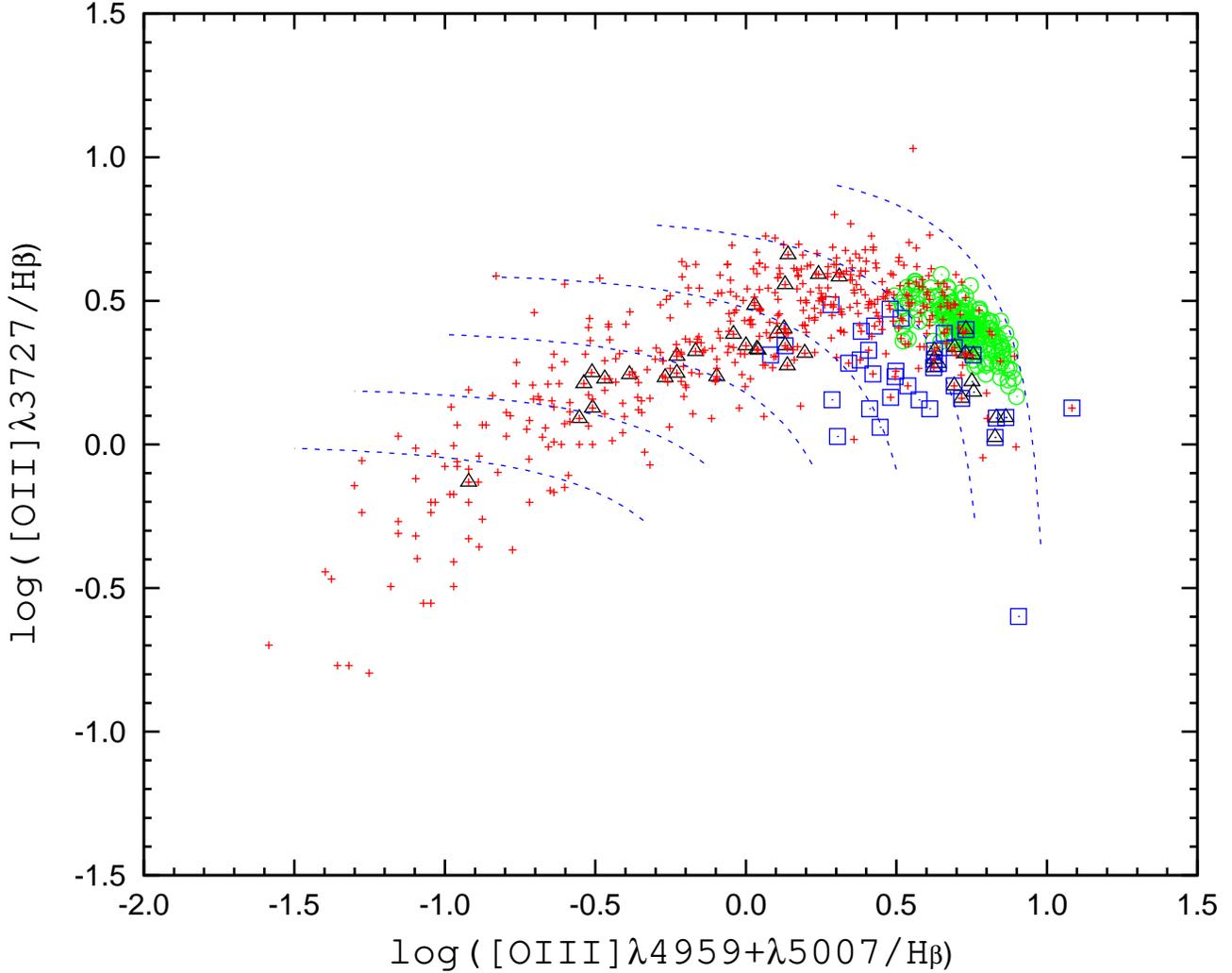}}
\caption{The [O\,{\sc iii}]$\lambda$4959+$\lambda$5007/H$\beta$ versus 
[O\,{\sc ii}]$\lambda$3727/H$\beta$ diagram. 
The open (green) circles are SDSS H\,{\sc ii} regions.
The open (blue) squares are H\,{\sc ii} regions in nearby galaxies with 
measured $t_{\rm 2,O}$ and $t_{\rm 3,O}$ temperatures. 
The open (black) triangles are H\,{\sc ii} regions in nearby galaxies 
used for deriving of the ff relation for 
$Q_{\rm 2,O}$ in \citet{pilyuginetal09}. 
The small (red) plus signs are H\,{\sc ii} regions in nearby galaxies 
\citep{pilyuginetal04}.
The short--dashed lines are lines with logR$_{23}$=0.0, 0.2, 0.4, 0.6, 0.8, 
1.0 from bottom to top (from left to right).
The figure is in colour in the on-line version of the paper.
}
\label{figure:r3r2}
\end{figure*}

Our final list includes only 
high-metallicity objects, those with 12+log(O/H) $\ga$ 8.2, 
which gives a total of 181 H\,{\sc ii} region spectra. 
(The oxygen abundances were estimated using the measured electron 
temperatures $t_{\rm 3,O}$ and  $t_{\rm 2,O}$.) 
Several H\,{\sc ii} regions have repeated observations 
(7 objects have 2 spectra, and 5 objects have 3 spectra)
so that the number of H\,{\sc ii} regions (164) is less than 
the number of spectra (181).
We have excluded 
low-metallicity H\,{\sc ii} regions from the present study 
for the following reason. 

A key part of our analysis rests on 
the comparison between two $t_{\rm 2,O}$--$t_{\rm 3,O}$ diagrams, one
 for SDSS H\,{\sc ii} regions and the other one 
for H\,{\sc ii} regions in nearby galaxies. 
In a previous study \citep{pilyuginetal09}, we have considered 
the $t_{\rm 2,O}$--$t_{\rm 3,O}$ diagram for high-metallicity H\,{\sc ii} 
(12+log(O/H) $\ga$ 8.2) regions in nearby galaxies. 
We have fitted the $t_{\rm 2,O}$--$t_{\rm 3,O}$ 
relation by the expression of the type 
\begin{equation}
t_{2} = a_0 + a_1\,t_{3}  .
\label{equation:mu}   
\end{equation}
The $t_{\rm 2,O}$--$t_{\rm 3,O}$ diagram exhibits some non-linearity.
This non-linearity may be explained in two ways:  
it is artificial and caused by the linear form 
adopted for the $ff$ relations for both [O\,{\sc iii}] and [O\,{\sc ii}] lines
(recall that the $ff$ relation, or the flux--flux relation, links the 
auroral  and nebular oxygen line fluxes in spectra of  
H\,{\sc ii} regions).  
Perhaps a more complex expression may give a better fit to the 
$ff$ relations. It may also be that the linear expression adopted for the 
$t_{\rm 2,O}$--$t_{\rm 3,O}$ relation is not a good approximation, and that 
an expression of the type, 
\begin{equation}
\frac{1}{t_{2}} = a_0 + a_1\frac{1}{t_{3}}  ,
\label{equation:on}   
\end{equation}
as suggested by \citet{pageletal92}, is more realistic. 
Thus, the relations derived by \citet{pilyuginetal09} 
may be considered as a first-order approximation.   
Extrapolation of the $t_{\rm 2,O}$--$t_{\rm 3,O}$ 
relation established for high-metallicity objects 
(12+log(O/H) $\ga$ 8.2) to the low-metallicity range 
may be unreliable. Therefore, only high-metallicity 
SDSS H\,{\sc ii} regions are considered here. 
The low-metallicity ones will be discussed elsewhere.

The wavelength range of the SDSS spectra is 3800 -- 9300\AA\ so that
for nearby galaxies with redshift $z$ $\la$ 0.02, the 
[O\,{\sc ii}]$\lambda$3727+$\lambda$3729 emission line is out of that range. 
The absence of this line prevents the determination of the 
oxygen abundance and hence   
the use of SDSS spectra of nearby galaxies in our study. 
Thus, the galaxies in our sample have redshifts ranging between  
$\sim$ 0.023 and $\sim$ 0.167,  
i.e. they are more distant than $\sim$ 100 Mpc.

For comparison with the SDSS H\,{\sc ii} region sample, we will be using the 
data for high-metallicity  (12+log(O/H) $\ga$ 8.2) single H\,{\sc ii} regions  
in nearby galaxies for which recent measurements of 
$t_{\rm 3}$ and $t_{\rm 2}$ are available. 
Such data have been compiled by \citet{pilyuginetal09}. 
We have added to this compilation the  recent spectrophotometric measurements of  
\citet{bresolinetal09a,bresolinetal09b,estebanetal09,savianeetal08}.

\subsection{General properties of the SDSS HII region sample}
%=====================

We now discuss some 
general characteristics of the SDSS H\,{\sc ii} region sample.

The intensities of strong, easily measured lines can be used to separate  
different types of emission-line objects according to their main  
excitation mechanism. \citet{baldwinetal81}  
have proposed a diagram where 
the excitation properties of H\,{\sc ii} regions are studied by 
plotting the low-excitation [N\,{\sc ii}]$\lambda$6584/H$\alpha$ 
line ratio against the high-excitation [O\,{\sc iii}]$\lambda$5007/H$\beta$ 
line ratio. 
Using this diagnostic diagram, we have excluded 
all objects above the solid line given by the equation
\begin{equation}
\log (\mbox{\rm [O\,{\sc iii}]$\lambda$5007/H$\beta$}) =
\frac{0.61}{\log (\mbox{\rm [N\,{\sc ii}]$\lambda$6584/H$\alpha$})-0.05} +1.3
\label{equation:kauff}   
\end{equation}
which separates objects with H\,{\sc ii} spectra from those 
containing an AGN \citep{kauffmannetal03}.
 The SDSS H\,{\sc ii} regions included in our final list  
are shown in upper panel of Fig.~\ref{figure:kauffmann} by open circles. 
For comparison, the lower panel of  Fig.~\ref{figure:kauffmann} shows 
a similar diagram for H\,{\sc ii} regions in nearby galaxies, as  
compiled by \citet{pilyuginetal04}.

Fig.~\ref{figure:ne} shows the density-sensitive  
[S\,{\sc ii}]$\lambda$6717/[S\,{\sc ii}]$\lambda$6731 line ratio 
as a function of the 
[S\,{\sc ii}]$\lambda$6717+$\lambda$6731 emission line fluxes
(upper panel) and of the $R_3$ line fluxes (lower panel). 
The expected zero density limit ([S\,{\sc ii}]$\lambda$6717/[S\,{\sc ii}]$\lambda$6731 = 1.44 
at $N_{\rm e}$=1 cm$^{-3}$ with $t_2$ = 1.0), is shown by the solid line. 
The dashed and dotted lines show respectively 
the line ratios corresponding to   
$N_{\rm e}$=100 cm$^{-3}$ ([S\,{\sc ii}]$\lambda$6717/[S\,{\sc ii}]$\lambda$6731 = 1.29) and 
$N_{\rm e}$=200 cm$^{-3}$ ([S\,{\sc ii}]$\lambda$6717/[S\,{\sc ii}]$\lambda$6731 = 1.18). 
All of the H\,{\sc ii} regions in our final sample  
have an electron density $N_{\rm e}$ $\leq$ 200 cm$^{-3}$, with   
the majority of them having $N_{\rm e}$ $\leq$ 100 cm$^{-3}$. 
They are thus all in the low-density regime, as 
is typical of the majority of extragalactic H\,{\sc ii} regions 
\citep{zkh,bresolinetal05}. 
A few SDSS H\,{\sc ii} regions in the original sample had 
$N_{\rm e}$ greater than 200 cm$^{-3}$, but they 
were excluded from the final sample so that  
the low-density approximation can be adopted 
for subsequent analysis. 
To determine the electron temperature, $N_{\rm e}$=50 cm$^{-3}$ 
has been adopted for all 
the objects in our final sample. 
In the electron temperature determination, we have assumed   
that the H\,{\sc ii} region is homogeneous and 
that there is no large difference between the electron 
densities in different ionic zones such as those of the  
S$^+$, O$^+$, and O$^{++}$ ions. 
If this is not the case, the electron temperatures estimated from  
$N_{\rm e}$(S$^+$) can have appreciable errors depending on the 
change of density across the nebula.

Fig.~\ref{figure:r3r2} shows the classical diagnostic diagram of 
[O\,{\sc iii}]$\lambda$4959+$\lambda$5007 versus 
[O\,{\sc ii}]$\lambda$3727. 
The open (green) circles represent SDSS H\,{\sc ii} regions, 
the open (blue) squares H\,{\sc ii} regions in nearby galaxies with 
measured $t_{\rm 2,O}$ and $t_{\rm 3,O}$ temperatures from the sample mentioned 
above, 
and the small (red) plus signs H\,{\sc ii} regions in nearby galaxies 
\citep{pilyuginetal04}.
The open (black) triangles show H\,{\sc ii} regions in nearby galaxies 
(with measured $t_{\rm 2,O}$ temperatures) used for deriving the $ff$ 
relation for $Q_{\rm 2,O}$ in \citet{pilyuginetal09}. 
Examination of Fig.~\ref{figure:r3r2} shows that 
the SDSS H\,{\sc ii} region points fall in the part of the diagram 
populated by the H\,{\sc ii} regions in nearby galaxies. 
This is in agreement with result of \citet{stasinskaizotov03}.

%====================================       Fig  No 4       log Lb -  N star O7 V
\begin{figure}
\resizebox{1.00\hsize}{!}{\includegraphics[angle=000]{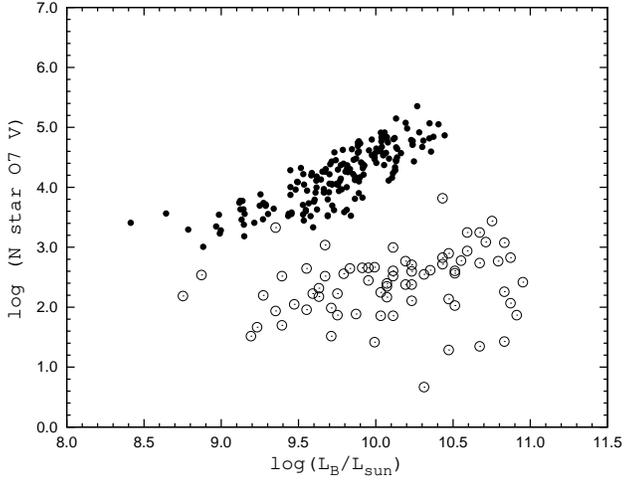}}
\caption{
The equivalent number of $O7~V$ stars  $N_{\rm O7~V}$ 
responsible for the excitation 
of H\,{\sc ii} regions versus the host galaxy luminosity $L_{\rm B}$ diagram.  
The filled circles represent SDSS objects.
The open circles show mean values of the three brightest H\,{\sc ii} 
regions in nearby galaxies \citep{kennicutt1988}. 
}
\label{figure:lbns}
\end{figure}

%====================================    Fig  No 5        t2 - t3 diagram observ
\begin{figure}
\resizebox{1.00\hsize}{!}{\includegraphics[angle=000]{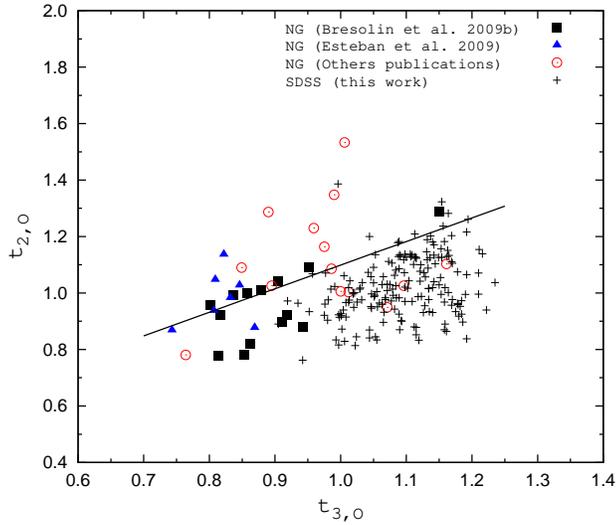}}
\caption{ 
The $t_{\rm 2,O}$ -- $t_{\rm 3,O}$ diagram. 
The (black) plus signs represent SDSS H\,{\sc ii} regions.
The H\,{\sc ii} regions in nearby galaxies are shown 
by filled (black) squares \citep{bresolinetal09b}, 
filled (blue) triangles \citep{estebanetal09}, 
and open (red) circles (data from other sources). 
The solid line shows the $t_{\rm 2,O}$--$t_{\rm 3,O}$ relation
derived for H\,{\sc ii} regions in nearby galaxies \citep{pilyuginetal09}.
The figure is in colour in the on-line version of the paper.
} 
\label{figure:ttobs}
\end{figure}

%====================================    Fig  No 6   T mean  - T individual
\begin{figure}
\resizebox{1.00\hsize}{!}{\includegraphics[angle=000]{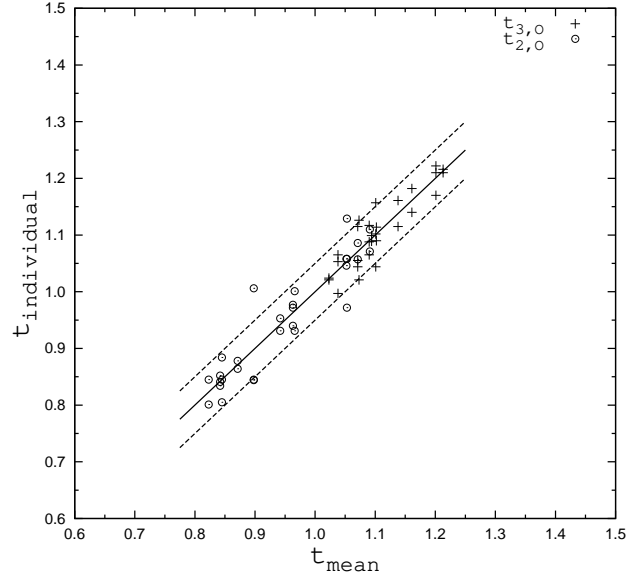}}
\caption{Individual electron temperatures versus  
average electron temperatures 
for SDSS H\,{\sc ii} regions in which electron temperatures 
have been measured from two or three different spectra. 
Plus signs show $t_{\rm 3,O}$ temperatures, while 
open circles show $t_{\rm 2,O}$ temperatures. 
The solid line is the locus of equal values, while the dashed lines represent 
shifts of $\pm$0.05 along the y-axis.
} 
\label{figure:tterr}
\end{figure}

%====================================    Fig  No 7       dt2 
\begin{figure}
\resizebox{1.00\hsize}{!}{\includegraphics[angle=000]{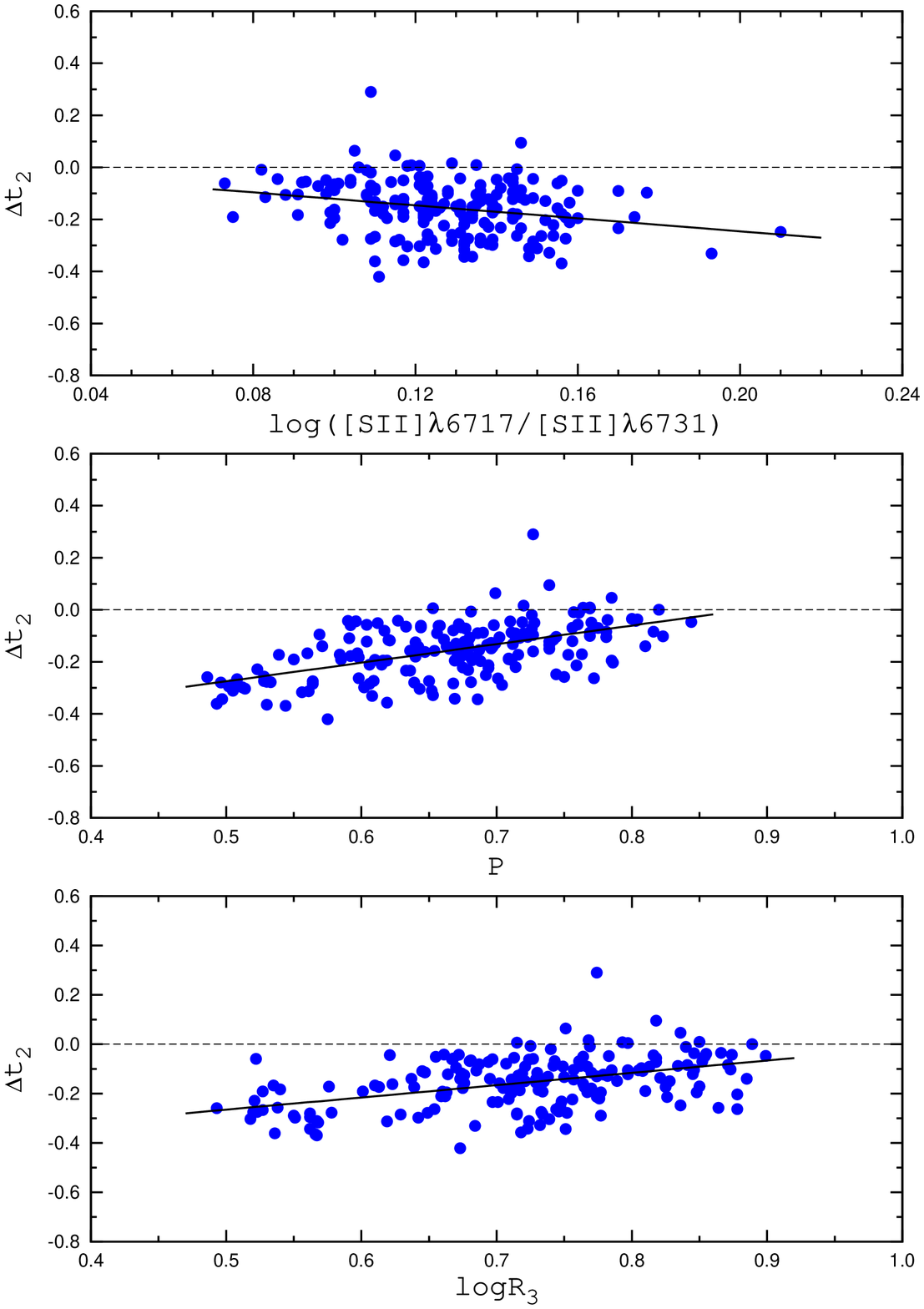}}
\caption{ 
Deviations of $t_{\rm 2,O}$ from the $t_{\rm 2,O}$--$t_{\rm 3,O}$ relation as a 
function of the density-sensitive 
[S\,{\sc ii}]$\lambda$6717/[S\,{\sc ii}]$\lambda$6731 
line ratio ({\it  upper panel}), 
of the excitation parameter $P$ ({\it middle panel}), and 
of the  $R_3$ line flux ({\it bottom panel}).
The solid lines show the obtained correlations, Table~\ref{table:dt}.
} 
\label{figure:dt2}
\end{figure}

An important characteristic of a H\,{\sc ii} region is the number of the ionizing stars it contains. 
Under the assumption of an ionization-bounded and dust-free nebula, the
H$\beta$ luminosity provides an estimate of the ionizing flux. 
The ionizing flux can be expressed in terms of the number of so-called 
equivalent $O$ stars of a given subtype responsible for producing the ionizing 
luminosity. The number of zero-age main sequense $O7~V$ stars, $N_{\rm O7~V}$, is 
usually used to specify the ionizing flux. It can be 
easily derived from the observed H$\beta$ luminosity and the Lyman continuum 
flux of an individual $O7~V$ star. 
The number of ionizing photons from an $O7~V$  star is taken to be  
$N_{\rm Lc}$ = 5.62$\times$10$^{48}$ s$^{-1}$ \citep{martinsetal2005}. 
The distances to SDSS galaxies are calculated using :  
\begin{equation}
d = \frac{cz}{H_o} ,
\label{equation:d}   
\end{equation}
where $d$ is the distance in Mpc,
$c$ the light velocity in km s$^{-1}$,
$z$ the redshift, and 
$H_{\rm o}$ the Hubble constant taken to be equal to 
 72 ($\pm$8) km s$^{-1}$ Mpc$^{-1}$ 
\citep{freedman2001}. 
In general, the H$\beta$ flux can be affected by underlying 
absorption and extinction. 
Underlying absorption is automatically taken into account since we fit   
the H$\beta$ line profile with two Gaussians for emission 
and absorption simultaneously. We do not correct the 
H${\beta}$ flux for 
extinction, so the derived numbers of ionizing stars are  
lower limits. 
The $B$ magnitude of the SDSS galaxy is obtained from  
the SDSS $m_{\rm g}$ and $m_{\rm r}$ magnitudes from the relation  
\begin{equation}
m_{\rm B} = m_{\rm g} + 0.42(m_{\rm g} - m_{\rm r}) +0.22 ,
\label{equation:mb}   
\end{equation}
derived from the data of \citet{fukugita1996}.
The SDSS galaxies in our sample 
are shown by filled circles in the $N_{\rm O7~V}$ vs. $L_{\rm B}$ diagram 
(Fig.~\ref{figure:lbns}). 
\citet{kennicutt1988} has given the mean H$\alpha$ fluxes for the three 
brightest H\,{\sc ii} regions in a sample of nearby galaxies. 
Those fluxes, converted to 
equivalent numbers of $O7~V$ stars are shown in the same figure 
by open circles. 
Fig.~\ref{figure:lbns} shows that 
the SDSS H\,{\sc ii} regions contain many more (up to $\sim$3 order of magnitude) 
ionizing stars, when compared to the brightest H\,{\sc ii} regions in 
nearby galaxies. This is not a surprising result. 
On the one hand, the SDSS H\,{\sc ii} regions suffer from a strong 
selection effect since, for distant galaxies, 
the weak [O\,{\sc iii}]$\lambda$4363 and 
[O\,{\sc ii}]$\lambda$7320+$\lambda$7330 auroral lines 
are detectable in only very bright 
H\,{\sc ii} regions. On the other hand,  
the SDSS spectra are obtained through 3-arcsec diameter fibers. 
The redshifts of our SDSS sample range from $\sim$0.025 to $\sim$0.17. 
At a redshift of $z$=0.025, 
the projected aperture diameter is $\sim$ 1.5 kpc, while it is  $\sim$ 10 kpc
at a redshift of $z$=0.17. 
This suggests that SDSS spectra are closer to global spectra of 
galaxies, i.e. composite nebulae including multiple star clusters, 
rather than to spectra of individual H\,{\sc ii} regions. 

\section{The ${\lowercase{t}}_2$ -- ${\lowercase{t}}_3$ diagram}
%=====================
 
%*************************** 
%\subsection{The $t_2$--$t_3$ diagram}
%*************************** 

To convert the quantities $Q_{\rm 3,O}$ and $Q_{\rm 2,O}$ to the electron 
temperatures $t_{\rm 3,O}$ and $t_{\rm 2,O}$, we have used the five-level-atom 
model, together with recent atomic data  for the O$^{+}$ and O$^{++}$ ions.
The Einstein coefficients of the 
spontaneous transitions for the five low-lying levels 
for both ions were taken from \cite{froese2004}.  The energy levels were taken 
from \citet{edlen1985} for O$^{++}$ and from \citet{wenaker1990} for O$^+$. 
The effective cross-sections, or effective collision strengths, for  
electron impact were taken from \citet{aggarwal1999} for O$^{++}$ and 
from \citet{pradhan2006} for O$^{+}$. 
We have fitted their 
tabulated data by a second-order polynomial in temperature. 
 The derived $t_{\rm 3,O}$ and $t_{\rm 2,O}$ 
are listed respectively in columns 11 and 12 of Table~\ref{table:list} 
in the Appendix,  
available on line.

Fig.~\ref{figure:ttobs} shows the $t_{\rm 2,O}$--$t_{\rm 3,O}$ diagram.
The SDSS sample H\,{\sc ii} regions are shown by (black) plus signs. 
For comparison, the data for high-metallicity H\,{\sc ii} regions 
in nearby galaxies with measured $t_{\rm 2,O}$ and $t_{\rm 3,O}$ temperatures
are also plotted. The data of  
\citet{bresolinetal09b} are shown 
by filled (black) squares, that of \citet{estebanetal09} by 
filled (blue) triangles, and that from other sources 
(discussed above) by open (red) circles. 
The solid line is the $t_{\rm 2,O}$--$t_{\rm 3,O}$ relation 
derived for  H\,{\sc ii} regions in nearby  galaxies 
\citep{pilyuginetal09}. It is of the form:
\begin{equation}
t_{\rm 2,O} = 0.835\,t_{\rm 3,O} + 0.264 
\label{equation:t3t209}   
\end{equation}
 Fig.~\ref{figure:ttobs} shows that some H\,{\sc ii} 
regions lie above that relation. 
It has been found \citep{pilyuginetal09} that H\,{\sc ii} regions with 
strong nebular $R_3$ line fluxes do not follow the 
$t_{\rm 2,O}$--$t_{\rm 3,O}$ relation: they are shifted towards higher 
$t_{\rm 2,O}$ temperatures. This shift can be caused by the fact that 
the low-lying metastable levels in some ions can be 
excited not only by the electron collisions, but also by 
recombination processes \citep{rubin86,liuetal00,stasinska05}. Since  
recombination-excited emission of [O\,{\sc ii}] in the auroral  
$\lambda$7325 radiation will  occur in the O$^{++}$ zone, the effect on the  
collision-excited lines will depend on the R$_3$ line flux. As  
a result, one may expect a significant shift toward higher values of 
$t_{\rm 2,O}$ for H\,{\sc ii} regions with strong $R_3$ line fluxes. 

Examination of Fig.~\ref{figure:ttobs} clearly shows that there is a large 
scatter of $t_{\rm 2,O}$ for a given $t_{\rm 3,O}$. Moreover, the majority  
of the SDSS H\,{\sc ii} regions lie below the $t_{\rm 2,O}$--$t_{\rm 3,O}$ 
relation established for H\,{\sc ii} regions in nearby  galaxies.
Although the SDSS H\,{\sc ii} regions have strong nebular 
$R_3$ line fluxes, they lie below, not above the solid line as  
the H\,{\sc ii} regions with strong nebular $R_3$ line fluxes in nearby 
galaxies. In conjunction with the lower $t_{\rm 2,O}$ temperatures,
the SDSS H\,{\sc ii} regions show also a shift 
towards higher $t_{\rm 3,O}$ temperatures.

Can uncertainties in the electron temperatures measurements be responsible 
for the observed scatter and shift of the SDSS H\,{\sc ii} regions in the 
$t_{\rm 2,O}$--$t_{\rm 3,O}$ diagram? 
We noted above that several SDSS H\,{\sc ii} regions 
possess two or three individual spectra with measurable 
temperature-sensitive nebular to auroral 
$Q_{\rm 2,O}$ and $Q_{\rm 3,O}$ line ratios.  
Comparison between separate measurements of the electron temperatures for 
the same H\,{\sc ii} region can give us an indication of their accuracies.
 In Fig.~\ref{figure:tterr}, we have plotted
 the individual value of the electron 
temperature versus the average value 
for SDSS H\,{\sc ii} regions that have electron temperatures 
measured separately from two or three spectra. 
The plus signs show $t_{\rm 3,O}$ temperatures, and  
the open circles $t_{\rm 2,O}$ temperatures. 
The solid line is the line of equal values and the dashed lines show  
shifts of $\pm$0.05 along the y-axis.
Inspection of Fig.~\ref{figure:tterr} shows that the typical scatter 
between individual measurements of both $t_{\rm 2,O}$ and 
$t_{\rm 3,O}$  is usually less than $\sim$ 1000~K. 
This can be considered as an estimate of the uncertainties in the electron 
temperature measurements. Then,
 these  
uncertainties are considerably less than 
the observed scatter and shift of the SDSS H\,{\sc ii} regions in the 
$t_{\rm 2,O}$--$t_{\rm 3,O}$ diagram.

%++++++++++++++++++ Table 1    Correlations
\begin{table}
\caption[]{\label{table:dt}
The coefficients a$_0$ and a$_1$ in the relations 
$\Delta$t$_2$=a$_0$+a$_1$X, Eq.(\ref{equation:cc}), 
and the values of the correlation coefficient.
}
\begin{center}
\begin{tabular}{lrrc} \hline \hline
X                                                        & 
a$_0$                                                    & 
a$_1$                                                    & 
correlation                                              \\
                                                         & 
                                                         & 
                                                         & 
coefficient                                              \\  \hline
log([S\,{\sc ii}]$\lambda$6717/[S\,{\sc ii}]$\lambda$6731) &   0.00 & --1.24 & --0.25   \\        
$P$                                                        & --0.63 &   0.71 &   0.55   \\        
log($R_3$)                                                 & --0.51 &   0.50 &   0.45   \\        
\hline
\hline 
\end{tabular}\\
\end{center}
\end{table}

We define the deviation $\Delta$$t_{\rm 2,O}$ 
of $t_{\rm 2,O}$ from the $t_{\rm 2,O}$--$t_{\rm 3,O}$  
relation as the difference between the measured $t_{\rm 2,O}$ and that 
obtained from the $t_{\rm 2,O}$--$t_{\rm 3,O}$ relation  
for a given $t_{\rm 3,O}$.  
Fig.~\ref{figure:dt2} shows $\Delta$$t_{\rm 2,O}$ =  
$t_{\rm 2,O}$--$t^r_{\rm 2,O}$ as a function of the   
density-sensitive 
[S\,{\sc ii}]$\lambda$6717/[S\,{\sc ii}]$\lambda$6731 line ratio (upper panel), 
of the excitation parameter $P$ (middle panel),  
and of the $R_3$ line flux (lower panel). 
The correlations of $\Delta$$t_{\rm 2,O}$ with different parameters 
were examined by obtaining relations of the type
\begin{eqnarray}
\Delta t_{\rm 2,O} = a_0 + a_1\;X
\label{equation:cc}
\end{eqnarray}
where $X$ is successively
log([S\,{\sc ii}]$\lambda$6717/[S\,{\sc ii}]$\lambda$6731), 
$P$, and  log($R_3$).
The derived coefficients a$_0$ and a$_1$ and the correlation 
coefficients are given in Table~\ref{table:dt}.
The derived relations are shown in Fig.~\ref{figure:dt2} by
solid lines.
It is seen that the deviations $\Delta$$t_{\rm 2,O}$ 
do not correlate strongly with  electron density, 
but do show some weak correlation (the correlation coefficient 
ia $\sim$0.5) with $P$ and $R_3$.

%*************************** 
\section{On a possible origin of the scatter in the 
 ${\lowercase{t}}_2$ -- ${\lowercase{t}}_3$ diagram} 
%*************************** 

We have remarked above that 
that the spectra of the SDSS H\,{\sc ii} regions correspond more 
to those of composite 
nebulae than to those of individual  H\,{\sc ii} regions.
Can this be responsible for the scatter and shift of the 
SDSS H\,{\sc ii} regions relative to the 
$t_{\rm 2,O}$--$t_{\rm 3,O}$ relation?

%====================================    Fig  No 8         Ercolano models
\begin{figure}
\resizebox{1.00\hsize}{!}{\includegraphics[angle=000]{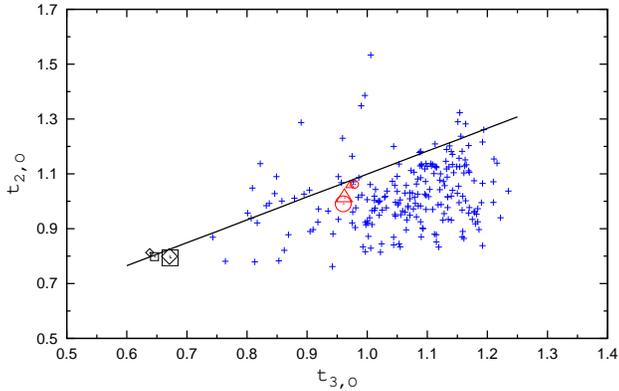}}
\caption{ 
The $t_{\rm 2,O}$--$t_{\rm 3,O}$ diagram for observed 
and modeled \citep{ercolanoetal07} H\,{\sc ii} regions. 
The SDSS and nearby  H\,{\sc ii} regions are 
shown by plus signs. 
The key for the model symbols is given in Table~\ref{table:legend}.
The solid line shows the $t_{\rm 2,O}$--$t_{\rm 3,O}$ relation
derived for H\,{\sc ii} regions in nearby galaxies \citep{pilyuginetal09}.
The figure is in colour in the on-line version of the paper.
} 
\label{figure:ercol}
\end{figure}

%++++++++++++++++++ Table 2    Legend
\begin{table}
\caption[]{\label{table:legend}
Key for the symbols representing the \citet{ercolanoetal07} models 
in Fig.~\ref{figure:ercol}. 
}
\begin{center}
\begin{tabular}{lcccc} \hline \hline
Symbol                                                   & 
star                                                     & 
geometry                                                 & 
Z/Z$_{\sun}$                                             & 
model                                                    \\ 
                                                         & 
distrib.                                                 & 
                                                         & 
                                                         & 
                                                         \\  \hline
small black rhomb   &  central  &  sphere  &  1.0  &  CSp1.0 \\  
large black rhomb   &  distrib. &  sphere  &  1.0  &  FSp1.0 \\  
small black square  &  central  &  shell   &  1.0  &  CSh1.0 \\  
large black square  &  distrib. &  shell   &  1.0  &  FSh1.0 \\  
small red circle    &  central  &  sphere  &  0.4  &  CSp0.4 \\  
large red circle    &  distrib. &  sphere  &  0.4  &  FSp0.4 \\  
small red triangle  &  central  &  shell   &  0.4  &  CSh0.4 \\  
large red triangle  &  distrib. &  shell   &  0.4  &  FSh0.4 \\  
\hline 
\end{tabular}\\
\end{center}
\end{table}

\citet{ercolanoetal07} have shown that the temperature structure of models 
with centrally concentrated ionizing sources may be quite different  
from those of models where the ionizing sources are 
randomly distributed within the volume, with generally non-overlapping 
Str\"omgren spheres.  
These differences may 
contribute to the scatter the $t_{\rm 2,O}$ -- t$_{\rm 3,O}$ diagram. 
The models of \citet{ercolanoetal07} 
are shown along with the SDSS and nearby galaxy samples 
in the $t_{\rm 2,O}$--$t_{\rm 3,O}$ diagram in Fig.~\ref{figure:ercol}. 
The key for the different model symbols is given in Table~\ref{table:legend}.
Examination of Fig.~\ref{figure:ercol} shows that the positions of 
the composite nebulae, consisting of several distributed H\,{\sc ii} regions, 
are indeed shifted lower 
relatively to the positions of the single nebulae. 
However, the predicted shift is considerably smaller 
than the observed shifts of the SDSS and nearby galaxy data points 
(plus signs).

%====================================    Fig  No 9     t2 - t3   mix of many objects
\begin{figure}
\resizebox{1.00\hsize}{!}{\includegraphics[angle=000]{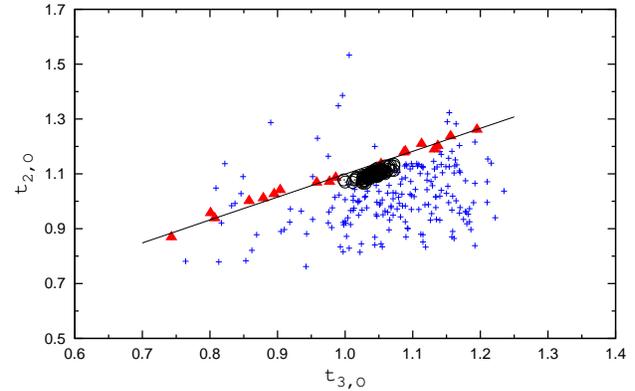}}
\caption{The t$_{\rm 2,O}$ -- t$_{\rm 3,O}$ diagram. 
Open (black) circles represent artificial composite H\,{\sc ii} regions 
consisting of randomly scaled components (shown by filled 
(red) triangles). 
Plus (blue) signs show SDSS and nearby H\,{\sc ii} regions. 
The continuous line is the $t_{\rm 2,O}$--$t_{\rm 3,O}$  relation \citep{pilyuginetal09}.
The figure is in colour in the on-line version of the paper.
}
\label{figure:ttmix}
\end{figure}

The composite nebula can be ionised by sources of 
different temperatures.  \citet{ercolanoetal07} have 
considered the case of an ionising set composed of two stellar 
populations, a 37 $M_{\sun}$ and a 56 $M_{\sun}$ population, with half of 
the ionising photon being emitted by each population. 
These masses correspond to stars of spectral types $O5$ and $O3$ respectively 
\citep{martinsetal2005}. 
However, those authors did not consider 
the case of a single stellar population. 
%a 37 $M_{\sun}$ and a 56 $M_{\sun}$ population, 
so that a direct comparison between the ionisation 
structure produced by single and composite populations is not possible. 
We have performed Monte Carlo runs to model  
 the temperature shifts from the $t_{\rm 2,O}$--$t_{\rm 3,O}$ 
relation caused by having several ionizing sources
with different temperatures within the same aperture in the following way.  
 We have assumed that each SDSS 
H\,{\sc ii} region consists of several components, with every component 
following the $t_{\rm 2,O}$--$t_{\rm 3,O}$ relation. These components 
are shown by filled (red) triangles in 
Fig.~\ref{figure:ttmix}.

If the H\,{\sc ii} region line fluxes are given on a scale where 
$F$(H${\beta}$)=1, then the total $F$(H${\beta}$) flux of the composite 
H\,{\sc ii} region is given by the expression: 
\begin{equation}
F({\rm H}{\beta})  =  \sum\limits_{{\rm j}=1}^{{\rm j}={\rm n}} w_{\rm j} 
\label{equation:hbran}   
\end{equation}
and the flux $F$(X$_{\lambda _{\rm k}}$) of the composite H\,{\sc ii} region  in the 
line X$_{\lambda _{\rm k}}$ is given by:
\begin{equation}
F({\rm X}_{\lambda _{\rm k}})  = \frac{\sum\limits_{{\rm j}=1}^{{\rm j}={\rm n}} 
w_{\rm j} F_{\rm j}({\rm X}_{\lambda _{\rm k}})}
                       {\sum\limits_{{\rm j}=1}^{{\rm j}={\rm n}} w_{\rm j}}
\label{equation:ran}   
\end{equation}
where $n$ is the number of components in the composite H\,{\sc ii} region, 
which we take to be equal to 18, and 
$w_{\rm j}$ are numbers between 0 and 1 produced by a random number generator. 

We have performed 100 Monte Carlo runs, producing  
in each run an artificial spectrum of composite H\,{\sc ii} 
regions.
For every computed spectrum, we have estimated the electron 
temperatures $t_{\rm 2,O}$ and t$_{\rm 3,O}$. 
The positions of those artificial 
composite H\,{\sc ii} regions in the $t_{\rm 2,O}$--$t_{\rm 3,O}$ diagram 
are shown by open (black) circles in Fig.~\ref{figure:ttmix}. 
The model points are shifted below the solid line, but again 
the shift is rather small.

%====================================    Fig No 10       t2 - t3  mix of 2 and 3
\begin{figure}
\resizebox{1.00\hsize}{!}{\includegraphics[angle=000]{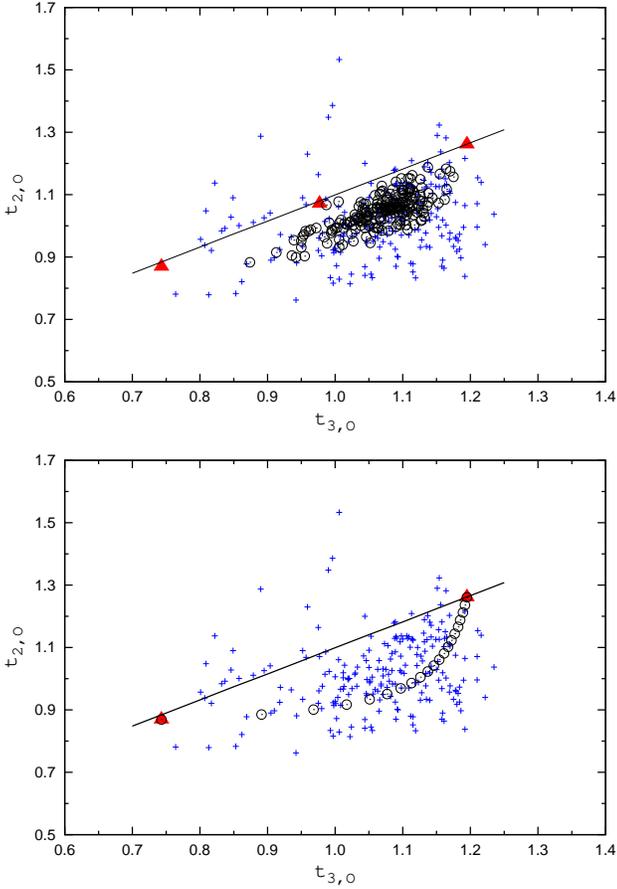}}
\caption{
The same as Fig.~\ref{figure:ttmix}, but for three-component ({\it top panel}) 
and two-component ({\it bottom panel}) composite H\,{\sc ii} regions. 
}
\label{figure:ttmix2}
\end{figure}

We next decrease the number of components to 3, with each 
component being widely separated in temperature from the others. 
They are shown by (red) triangles in the top panel of
 Fig.~\ref{figure:ttmix2}.   
We have computed 200 artificial spectra of composite H\,{\sc ii} 
regions. Like before, the contribution of each component 
is randomly scaled. 
The positions of those artificial spectra in the 
$t_{\rm 2,O}$--$t_{\rm 3,O}$ diagram are shown by 
the open (black) circles in the top panel of Fig.~\ref{figure:ttmix2}. 
It is seen that, in this case, there are composite H\,{\sc ii} 
regions that are significantly shifted below the $t_{\rm 2,O}$--$t_{\rm 3,O}$ 
relation, although they still cannot match the positions of the 
SDSS H\,{\sc ii} regions with the largest shifts.

We next compute artificial spectra of   
two-component H\,{\sc ii} regions. These are shown in the 
$t_{\rm 2,O}$--$t_{\rm 3,O}$ diagram by 
open (black) circles in the bottom panel of Fig.~\ref{figure:ttmix2}. 
In this case, $w_1$ and $w_2$ are not random numbers, but are 
chosen so that $w_1$ varies from 0 to 1 with 
a step of 0.05 and $w_2$ = 1--$w_1$. 
Again, those composite H\,{\sc ii} regions are significantly shifted from the 
$t_{\rm 2,O}$--$t_{\rm 3,O}$ relation, 
although they again do not reproduce the positions 
of the SDSS objects with the largest shifts.  
However, examination of Fig.~\ref{figure:ttmix2}  
suggests that the SDSS objects with 
the largest shifts, if real,  
can be reproduced by a composite object where the 
hot component has $t_{\rm 2,O}$ $\sim$ 1.3 and the
cool component $t_{\rm 2,O}$ $\sim$ 0.8. 
Unfortunately, we cannot test directly 
this expectation since the data for an actual  
H\,{\sc ii} region with both $t_{\rm 2,O}$ and $t_{\rm 3,O}$ measured, 
which follows the $t_{\rm 2,O}$--$t_{\rm 3,O}$ relation, and   
with $t_{\rm 2,O}$ $\sim$ 0.8 is not available.

%====================================    Fig No 11       z - dt2 
\begin{figure}
\resizebox{1.00\hsize}{!}{\includegraphics[angle=000]{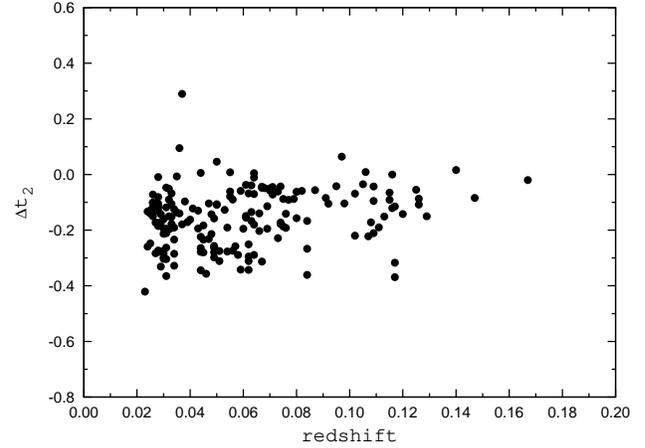}}
\caption{ 
Deviations of $t_{\rm 2,O}$ from the $t_{\rm 2,O}$--$t_{\rm 3,O}$ relation as a 
function of redshift. 
} 
\label{figure:zdt2}
\end{figure}

In summary, 
the observed $t_{\rm 2,O}$--$t_{\rm 3,O}$ diagram can be understood if 
the SDSS H\,{\sc ii} regions are composites of a few  H\,{\sc ii} regions with 
different emission line properties. The evolution of H\,{\sc ii} regions 
associated with stars and star clusters has been considered in many papers 
\citep[][among others]{stasinska78,stasinska80,mrs,dopitaevans86,
moyetal2001,stasinskaizotov03,dopitaetal06}. 
It is well established that the H\,{\sc ii} region spectral sequence depends 
on three fundamental parameters: the metallicity, the ionizing star cluster 
spectral energy distribution and the ionization parameter. 
\citet{ercolanoetal07} have shown that variations in the ionization 
parameter cannot be the main reason for the scatter in the 
$t_{\rm 2,O}$--$t_{\rm 3,O}$ diagram. Variations of 
the ionizing star cluster spectral energy distribution are governed by 
two independent parameters: age and metallicity. 
Then, the shifted SDSS H\,{\sc ii} regions should consist of components of 
different ages or/and different metallicities. 

Of these two parameters, which is mainly responsible 
for the differences in the emission line 
properties of the components of the SDSS H\,{\sc ii} regions?
Because of radial abundance gradients in the disks of spiral 
galaxies, the oxygen abundances in H\,{\sc ii} regions near the center of 
a galaxy and those at the periphery can differ by 
up to an order of magnitude \citep{pilyuginetal04}.  
We noted above that the SDSS spectra are obtained through 3-arcsec diameter 
fibers, and the projected aperture diameter increases with redshift, up to 
$\sim$ 10 kpc at the redshift of $z=0.17$. 
The range of galactocentric distances and, consequently, the range of 
metallicities of H\,{\sc ii} regions which contribute to the 
SDSS spectra, is larger for distant than for nearby galaxies. 
If metallicity variations are responsible for the differences in the 
emission line properties of the H\,{\sc ii} regions which 
contribute to the SDSS spectra, 
then the deviations of these H\,{\sc ii} regions 
from the $t_{\rm 2,O}$--$t_{\rm 3,O}$ 
relation should increase with redshift. 
In Fig.~\ref{figure:zdt2}, we plot the deviations $\Delta$$t_{\rm 2,O}$ 
of the SDSS objects as a function of redshift. 
There is not a significant correlation.
 
This suggests that age is the main factor 
of the scatter of the SDSS 
H\,{\sc ii} regions rather than metallicity.
It should be noted, however, that the evolution of H\,{\sc ii} 
regions associated with star clusters is generally 
accompanied by a variation of its 
chemical composition through a self-enrichment process
\citep{kunthsargent86,pilyugin92,pilyugin93}. Then young and old 
H\,{\sc ii} regions can have different 
chemical compositions. However these chemical differences are not 
seen in optical spectra of H\,{\sc ii} regions that are only  
several million years old. 
Thus, we conclude that  
it is the differences in the ionizing star cluster spectral energy 
distributions, caused by their age variations,
 that are responsible for the 
differences in the emission-line properties of 
the H\,{\sc ii} regions in SDSS galaxies.

Our explanation of the observed scatter and shift of the SDSS 
H\,{\sc ii} regions in the $t_{\rm 2,O}$--$t_{\rm 3,O}$ diagram is based on 
the hypothesis that SDSS spectra result from a mix of H\,{\sc ii} regions 
ionized by sources of different temperatures. This hypothesis is in  
line with results derived from considerations of other diagrams. 
\citet{stasinskaizotov03} have noted that the classical diagnostic diagram 
[O\,{\sc iii}]$\lambda$4959+$\lambda$5007 versus 
[O\,{\sc ii}]$\lambda$3727 is not completely understood 
in terms of pure photoionization models.  
\citet{stasinskaetal01} have proposed an 
additional heating mechanism to account for the largest  
[O\,{\sc ii}]$\lambda$3727 observed fluxes.
\citet{moyetal2001} have concluded that the best fits of emission-line ratios 
in starburst and H\,{\sc ii} galaxies 
are obtained with a combination of a high- and low-ionization components.
In terms of the present study, those conclusions mean that giant 
H\,{\sc ii} objects with the larges
[O\,{\sc ii}]$\lambda$3727 fluxes are 
composite H\,{\sc ii} regions. 

\citet{dopitaetal06} have 
modeled a H\,{\sc ii} region associated 
with star clusters where stars are being formed continuously. 
They have computed line flux-averaged spectra along evolutionary 
tracks of H\,{\sc ii} regions, and have found  
that the older evolved H\,{\sc ii} regions make a significant contribution 
to the low-ionization lines (e.g. [O\,{\sc ii}]$\lambda$3727),
 while high-ionization lines, 
such as [O\,{\sc iii}]$\lambda$4959+$\lambda$5007, will be mostly 
produced in the very youngest H\,{\sc ii} regions. 
In the light of our present results, one can expect the following 
picture for the models of \citet{dopitaetal06}. Continuous star formation 
can be considered as a sequence of many substarbursts of different ages, i.e. 
as a sequence of many ionizing sources with different spectral energy 
distributions. 
If all the substarbursts have a similar amplitude, i.e. if there are no 
large variations in the star formation rate,  
then the associated H\,{\sc ii} regions will lie close 
to the  $t_{\rm 2,O}$--$t_{\rm 3,O}$ relation (Fig.~\ref{figure:ttmix}). 
On the other hand, if 
a few substarbursts have a significantly higher amplitude, i.e. if 
the star formation rate varies appreciably, 
then the associated H\,{\sc ii} regions will show 
significant deviations from the  $t_{\rm 2,O}$--$t_{\rm 3,O}$ 
relation (Fig.~\ref{figure:ttmix2}).

To summarize, the presence of low-ionization components in giant H\,{\sc ii} 
regions appears to be substantiated by a number of different investigations. 
However, their origin is not yet clear. 
Are they old and cold H\,{\sc ii} regions as suggested by 
\citet{dopitaetal06}, 
or are they the ``low-ionization components excited by an additional 
heating mechanism'' as suggested by \citet{stasinskaetal01}, 
or are they the diffuse ionized medium? 
A detailed study of very cold H\,{\sc ii} regions can clarify this 
matter.

\section{Conclusions}
%=====================

Spectra of H\,{\sc ii} regions where   the two auroral 
[O\,{\sc iii}]$\lambda$4363 and [O\,{\sc ii}]$\lambda$7325 lines  
are detectable have been visually extracted from the Data Release 6 of the 
Sloan Digital Sky Survey (SDSS). Our final list consists of 181 spectra 
of high-metallicity (12+log(O/H) $\ga$ 8.2) SDSS H\,{\sc ii} regions.
The redshifts of our SDSS sample lie in the range from $\sim$0.025 to 
$\sim$0.17. 
The SDSS H\,{\sc ii} regions fall in the region 
in the  classical 
[O\,{\sc iii}]$\lambda$4959+$\lambda$5007 versus 
[O\,{\sc ii}]$\lambda$3727 
diagnostic diagram populated by the 
H\,{\sc ii} regions of nearby galaxies. 
At the same time, 
the number of equivalent $O7~V$ stars in the 
H\,{\sc ii} regions of the SDSS sample is considerably larger 
(up to three orders of magnitude) 
than that in the brightest H\,{\sc ii} regions in nearby galaxies.

The $t_{\rm 2,O}$--$t_{\rm 3,O}$ diagram for 
the SDSS sample is examined. It is found that 
there is a large scatter of the $t_{\rm 2,O}$ electron temperature,
for a given $t_{\rm 3,O}$ electron temperature. 
The majority of the SDSS H\,{\sc ii} regions lie below the 
$t_{\rm 2,O}$--$t_{\rm 3,O}$ relation obtained for  
H\,{\sc ii} regions in nearby galaxies, i.e. they
show a systematic shift towards lower $t_{\rm 2,O}$ 
temperatures or/and towards higher  $t_{\rm 3,O}$ temperatures. 

The observed scatter and shift of the SDSS H\,{\sc ii} 
regions in the $t_{\rm 2,O}$--$t_{\rm 3,O}$ diagram can be explained if 
the SDSS spectra correspond to a mix of 
H\,{\sc ii} regions ionized by sources of different temperatures. 
The sources of ionization can be centrally concentrated or distributed within 
the H\,{\sc ii} 
region volume. In the latter case,
 there is an additional scatter as found 
by \citet{ercolanoetal07}. 

%===========================================================================
\section*{Acknowledgments}

We are grateful to the referee of this paper for his/her constructive comments.
L.S.P. thanks the staff of the Instituto de Astrof\'{\i}sica de Andaluc\'{\i}a 
for hospitality during a visit when a significant part of this investigation 
was carried out. T.X.T thanks the financial support of NSF and NASA.

\appendix

\section{Online material. Table A1.}
%=================

Table~\ref{table:list} contains the dereddened line intensities and 
the electron temperatures of the SDSS H\,{\sc ii} regions.
The line intensities are given on a scale in which H$\beta$
=1.
The objects are listed in order of right ascension. 
The first column gives the order number of the object. 
The J2000.0 right ascension and declination  of each object 
are given in columns 2 and 3 respectively. They constitute the 
SDSS identificator 
of each object. Units of right ascension are hours, minutes, and seconds and 
units of declination are degrees, arcminutes, and arcseconds. 
The SDSS spectrum number, composed of the 
plate number, the modified Julian date of observations and the  
number of the fiber on the plate, is given in column 4.
The measured 
[O\,{\sc ii}]$\lambda$3727+$\lambda$3729,
[O\,{\sc iii}]$\lambda$4363, 
[O\,{\sc iii}]$\lambda$4959+$\lambda$5007, 
[S\,{\sc ii}]$\lambda$6717, 
[S\,{\sc ii}]$\lambda$6731, 
[O\,{\sc ii}]$\lambda$7320+$\lambda$7330 
line intensities are listed in columns from 5 to 10 respectively.
The derived electron temperatures t$_{\rm 3,O}$ and t$_{\rm 2,O}$ 
are given in columns 11 and 12 respectively. 

%++++++++++++++++++ Table 1    Lines and temperatures in the SDSS sample
\setcounter{table}{0}
\begin{table*}
%\rotate
%\tabletypesize{\tiny}
%\tabletypesize{\scriptsize}
%\tabletypesize{\small}
%\tabletypesize{\normalsize}
\caption[]{\label{table:list}
Dereddened line intensities and electron temperatures 
of a sample of HII regions in SDSS galaxies. 
The line intensities are given on a scale in which H$\beta$=1.
}
%\begin{flushleft}
\begin{center}
\begin{tabular}{rrrrrrcccccc} \hline \hline
No$^a$                                                   & 
RA$^b$                                                   & 
DEC$^c$                                                  & 
Spectrum$^d$                                             & 
[O\,{\sc ii}]                                            & 
[O\,{\sc iii}]                                           & 
[O\,{\sc iii}]                                           & 
[S\,{\sc ii}]                                            & 
[S\,{\sc ii}]                                            & 
[O\,{\sc ii}]                                            & 
t$_{\rm 3,O}$                                            & 
t$_{\rm 2,O}$                                            \\
                                                         & 
                                                         & 
                                                         & 
number                                                   & 
$\lambda$3727                                            & 
$\lambda$4363                                            & 
$\lambda$4959+$\lambda$5007                              & 
$\lambda$6717                                            & 
$\lambda$6731                                            & 
$\lambda$7325                                            & 
                                                         & 
                                                         \\   \hline
   1 &  00 06 57.01 &  00 51 26.0 &    686 52519   406 &  1.840 &  0.0619 &  7.477 &  0.173 &  0.128 &  0.0478 &  1.192 &  1.216 \\  
   2 &  00 20 12.06 & -00 43 06.3 &   1088 52929   183 &  2.675 &  0.0283 &  5.538 &  0.230 &  0.174 &  0.0537 &  1.014 &  1.043 \\  
   3 &  00 51 18.38 &  00 32 57.7 &    394 51913   469 &  3.530 &  0.0262 &  3.432 &  0.303 &  0.235 &  0.0493 &  1.158 &  0.870 \\  
   4 &  00 51 18.38 &  00 32 57.7 &    691 52199   626 &  2.693 &  0.0226 &  3.425 &  0.303 &  0.227 &  0.0516 &  1.102 &  1.017 \\  
   5 &  00 51 18.38 &  00 32 57.7 &    394 51812   466 &  3.262 &  0.0189 &  3.367 &  0.295 &  0.229 &  0.0454 &  1.044 &  0.869 \\  
   6 &  00 52 33.72 &  00 19 52.6 &    394 51913   493 &  3.609 &  0.0204 &  3.645 &  0.346 &  0.252 &  0.0467 &  1.044 &  0.841 \\  
   7 &  00 52 33.72 &  00 19 52.6 &    394 51876   494 &  3.690 &  0.0250 &  3.650 &  0.361 &  0.265 &  0.0491 &  1.115 &  0.852 \\  
   8 &  00 52 33.72 &  00 19 52.6 &    394 51812   487 &  3.617 &  0.0213 &  3.688 &  0.356 &  0.252 &  0.0459 &  1.055 &  0.834 \\  
   9 &  01 32 58.54 & -08 53 37.8 &    662 52147   466 &  2.725 &  0.0227 &  4.582 &  0.356 &  0.256 &  0.0564 &  1.005 &  1.061 \\  
  10 &  02 11 59.94 & -01 13 02.8 &    405 51816   365 &  3.481 &  0.0389 &  4.708 &  0.302 &  0.234 &  0.0447 &  1.192 &  0.838 \\  
  11 &  02 53 25.28 & -00 13 56.6 &    409 51871   033 &  2.690 &  0.0425 &  5.416 &  0.210 &  0.156 &  0.0458 &  1.169 &  0.957 \\  
  12 &  03 26 14.50 & -00 12 10.8 &    712 52179   156 &  3.114 &  0.0173 &  3.296 &  0.313 &  0.237 &  0.0373 &  1.022 &  0.814 \\  
  13 &  03 26 14.50 & -00 12 10.8 &    712 52199   158 &  2.980 &  0.0176 &  3.331 &  0.303 &  0.223 &  0.0390 &  1.024 &  0.845 \\  
  14 &  03 27 50.15 &  01 01 35.0 &    414 51869   524 &  3.268 &  0.0299 &  4.700 &  0.326 &  0.234 &  0.0753 &  1.088 &  1.129 \\  
  15 &  03 27 50.15 &  01 01 35.0 &    414 51901   546 &  2.990 &  0.0316 &  4.592 &  0.338 &  0.255 &  0.0541 &  1.118 &  0.987 \\  
  16 &  03 27 50.15 &  01 01 35.0 &    805 52586   550 &  3.538 &  0.0278 &  4.674 &  0.344 &  0.247 &  0.0728 &  1.065 &  1.058 \\  
  17 &  03 33 14.42 &  00 24 37.2 &    713 52178   481 &  3.711 &  0.0176 &  3.647 &  0.329 &  0.248 &  0.0447 &  0.997 &  0.816 \\  
  18 &  03 33 14.42 &  00 24 37.2 &    415 51810   486 &  3.397 &  0.0205 &  3.559 &  0.328 &  0.238 &  0.0444 &  1.053 &  0.845 \\  
  19 &  03 33 14.42 &  00 24 37.2 &    415 51879   497 &  3.093 &  0.0206 &  3.454 &  0.340 &  0.256 &  0.0460 &  1.065 &  0.896 \\  
  20 &  07 29 30.29 &  39 49 41.6 &   1733 53047   528 &  2.590 &  0.0364 &  5.778 &  0.229 &  0.182 &  0.0554 &  1.084 &  1.081 \\  
  21 &  07 48 06.30 &  19 31 46.9 &   1582 52939   335 &  2.381 &  0.0274 &  5.380 &  0.197 &  0.148 &  0.0421 &  1.013 &  0.975 \\  
  22 &  07 49 15.48 &  22 53 42.1 &   1203 52669   570 &  3.214 &  0.0426 &  5.222 &  0.279 &  0.213 &  0.0479 &  1.186 &  0.897 \\  
  23 &  07 53 32.81 &  46 44 39.6 &   1737 53055   555 &  2.602 &  0.0474 &  6.050 &  0.295 &  0.208 &  0.0583 &  1.169 &  1.111 \\  
  24 &  08 00 00.69 &  27 46 42.0 &    859 52317   070 &  2.197 &  0.0532 &  7.076 &  0.192 &  0.141 &  0.0450 &  1.152 &  1.055 \\  
  25 &  08 06 19.49 &  19 49 27.3 &   1922 53315   588 &  2.167 &  0.0331 &  5.801 &  0.208 &  0.159 &  0.0471 &  1.050 &  1.091 \\  
  26 &  08 08 24.94 &  23 08 40.8 &   1584 52943   372 &  2.985 &  0.0272 &  5.352 &  0.245 &  0.188 &  0.0519 &  1.012 &  0.967 \\  
  27 &  08 14 20.78 &  57 50 08.0 &   1872 53386   526 &  2.532 &  0.0287 &  5.091 &  0.285 &  0.206 &  0.0521 &  1.046 &  1.058 \\  
  28 &  08 20 09.11 &  23 39 19.6 &   1926 53317   369 &  2.655 &  0.0284 &  5.263 &  0.216 &  0.160 &  0.0470 &  1.032 &  0.976 \\  
  29 &  08 23 54.96 &  28 06 21.6 &   1267 52932   384 &  1.964 &  0.0358 &  5.716 &  0.186 &  0.145 &  0.0408 &  1.082 &  1.063 \\  
  30 &  08 24 53.13 &  39 06 39.1 &    894 52615   185 &  2.192 &  0.0535 &  6.689 &  0.186 &  0.142 &  0.0464 &  1.178 &  1.075 \\  
  31 &  08 25 27.65 &  29 57 39.2 &   1207 52672   506 &  3.054 &  0.0236 &  4.417 &  0.357 &  0.261 &  0.0582 &  1.029 &  1.014 \\  
  32 &  08 25 59.30 &  21 35 46.9 &   1927 53321   560 &  2.873 &  0.0331 &  4.605 &  0.361 &  0.250 &  0.0549 &  1.134 &  1.016 \\  
  33 &  08 26 30.79 &  22 14 24.0 &   1927 53321   570 &  2.941 &  0.0271 &  4.351 &  0.357 &  0.259 &  0.0537 &  1.080 &  0.992 \\  
  34 &  08 31 56.33 &  43 36 12.7 &    762 52232   224 &  2.873 &  0.0418 &  6.007 &  0.251 &  0.189 &  0.0610 &  1.121 &  1.077 \\  
  35 &  08 33 26.07 &  21 51 46.3 &   1929 53349   100 &  2.348 &  0.0287 &  4.958 &  0.249 &  0.187 &  0.0440 &  1.055 &  1.005 \\  
  36 &  08 33 43.89 &  43 19 51.7 &    762 52232   217 &  3.019 &  0.0274 &  4.612 &  0.337 &  0.242 &  0.0592 &  1.063 &  1.030 \\  
  37 &  08 37 11.33 &  23 11 58.1 &   1929 53349   593 &  2.731 &  0.0483 &  5.846 &  0.282 &  0.208 &  0.0569 &  1.192 &  1.065 \\  
  38 &  08 38 43.63 &  38 53 50.5 &    828 52317   148 &  1.946 &  0.0535 &  6.927 &  0.188 &  0.142 &  0.0463 &  1.163 &  1.151 \\  
  39 &  08 44 14.22 &  02 26 21.2 &    564 52224   216 &  1.680 &  0.0545 &  7.430 &  0.137 &  0.109 &  0.0390 &  1.142 &  1.134 \\  
  40 &  08 45 27.60 &  53 08 52.9 &    447 51877   361 &  2.231 &  0.0621 &  7.546 &  0.162 &  0.116 &  0.0410 &  1.190 &  0.995 \\  
  41 &  08 45 47.17 &  27 54 24.5 &   1588 52965   073 &  2.829 &  0.0421 &  5.301 &  0.284 &  0.202 &  0.0459 &  1.175 &  0.934 \\  
  42 &  08 56 17.61 &  26 40 12.7 &   1933 53381   460 &  2.347 &  0.0396 &  5.852 &  0.250 &  0.169 &  0.0518 &  1.111 &  1.101 \\  
  43 &  09 04 03.59 &  36 39 14.2 &   1212 52703   388 &  2.512 &  0.0479 &  5.978 &  0.238 &  0.174 &  0.0430 &  1.179 &  0.959 \\  
  44 &  09 04 18.11 &  26 01 06.3 &   1935 53387   204 &  2.691 &  0.0545 &  7.022 &  0.195 &  0.158 &  0.0624 &  1.165 &  1.133 \\  
  45 &  09 07 04.89 &  53 26 56.6 &    553 51999   342 &  2.182 &  0.0380 &  5.813 &  0.223 &  0.162 &  0.0421 &  1.098 &  1.021 \\  
  46 &  09 08 00.08 &  50 39 10.0 &    552 51992   027 &  2.934 &  0.0421 &  5.191 &  0.286 &  0.215 &  0.0516 &  1.184 &  0.973 \\  
  47 &  09 12 52.31 &  50 26 21.3 &    766 52247   434 &  2.868 &  0.0302 &  4.522 &  0.341 &  0.238 &  0.0669 &  1.106 &  1.137 \\  
  48 &  09 14 00.36 &  49 59 30.9 &    766 52247   225 &  3.272 &  0.0247 &  3.994 &  0.360 &  0.241 &  0.0575 &  1.077 &  0.972 \\  
  49 &  09 18 20.53 &  49 06 34.9 &    900 52637   388 &  2.313 &  0.0358 &  5.338 &  0.248 &  0.178 &  0.0533 &  1.107 &  1.129 \\  
  50 &  09 20 33.22 &  31 36 35.5 &   1938 53379   406 &  2.531 &  0.0464 &  6.494 &  0.246 &  0.183 &  0.0574 &  1.132 &  1.118 \\  
  51 &  09 28 06.23 &  38 07 57.0 &   1214 52731   142 &  1.711 &  0.0550 &  7.018 &  0.185 &  0.140 &  0.0437 &  1.169 &  1.203 \\  
  52 &  09 32 48.77 &  58 25 30.6 &    452 51911   487 &  2.483 &  0.0323 &  5.143 &  0.265 &  0.196 &  0.0483 &  1.084 &  1.026 \\  
  53 &  09 38 13.49 &  54 28 25.0 &    556 51991   224 &  1.865 &  0.0394 &  5.764 &  0.216 &  0.163 &  0.0428 &  1.115 &  1.126 \\  
  54 &  09 51 15.96 &  36 40 31.1 &   1596 52998   311 &  2.951 &  0.0253 &  3.701 &  0.358 &  0.264 &  0.0420 &  1.115 &  0.878 \\  
  55 &  09 51 15.96 &  36 40 31.1 &   1595 52999   532 &  3.094 &  0.0284 &  3.691 &  0.367 &  0.256 &  0.0425 &  1.161 &  0.864 \\  
  56 &  09 54 08.41 &  54 46 07.0 &    769 52282   575 &  2.648 &  0.0412 &  6.452 &  0.221 &  0.169 &  0.0475 &  1.089 &  0.983 \\  
  57 &  09 57 21.27 &  36 52 44.7 &   1596 52998   479 &  2.389 &  0.0169 &  3.368 &  0.195 &  0.164 &  0.0371 &  1.008 &  0.915 \\  
  58 &  10 03 07.78 &  13 03 26.0 &   1744 53055   450 &  3.053 &  0.0288 &  4.653 &  0.287 &  0.232 &  0.0679 &  1.078 &  1.106 \\  
  59 &  10 09 50.24 &  11 04 39.3 &   1597 52999   358 &  2.146 &  0.0329 &  5.216 &  0.251 &  0.189 &  0.0385 &  1.085 &  0.983 \\  
  60 &  10 11 39.16 &  10 10 42.2 &   1597 52999   251 &  2.686 &  0.0373 &  6.153 &  0.263 &  0.198 &  0.0507 &  1.071 &  1.009 \\  
  61 &  10 12 42.97 &  61 33 02.7 &    771 52370   490 &  2.340 &  0.0520 &  6.619 &  0.203 &  0.141 &  0.0520 &  1.170 &  1.105 \\  
  62 &  10 12 57.26 &  39 28 50.2 &   1357 53034   202 &  1.600 &  0.0563 &  7.461 &  0.158 &  0.126 &  0.0367 &  1.154 &  1.126 \\  
\hline 
\end{tabular}\\
\end{center}
\end{table*}

%+++++++++++++++++++++++++++++++++ Table 1    Lines in the SDSS sample
\setcounter{table}{0}
\begin{table*}
%\rotate
%\tabletypesize{\tiny}
%\tabletypesize{\scriptsize}
%\tabletypesize{\small}
%\tabletypesize{\normalsize}
\caption[]{
Continued.
}
%\begin{flushleft}
\begin{center}
\begin{tabular}{rrrrrrcccccc} \hline \hline
No$^a$                                                   & 
RA$^b$                                                   & 
DEC$^c$                                                  & 
Spectrum$^d$                                             & 
[O\,{\sc ii}]                                            & 
[O\,{\sc iii}]                                           & 
[O\,{\sc iii}]                                           & 
[S\,{\sc ii}]                                            & 
[S\,{\sc ii}]                                            & 
[O\,{\sc ii}]                                            & 
t$_{\rm 3,O}$                                            & 
t$_{\rm 2,O}$                                            \\
                                                         & 
                                                         & 
                                                         & 
number                                                   & 
$\lambda$3727                                            & 
$\lambda$4363                                            & 
$\lambda$4959+$\lambda$5007                              & 
$\lambda$6717                                            & 
$\lambda$6731                                            & 
$\lambda$7325                                            & 
                                                         & 
                                                         \\   \hline
  63 &  10 21 06.34 &  36 04 08.7 &   1957 53415   258 &  2.484 &  0.0372 &  5.072 &  0.299 &  0.209 &  0.0496 &  1.142 &  1.040 \\  
  64 &  10 28 47.65 &  39 49 41.5 &   1428 52998   568 &  2.808 &  0.0265 &  4.567 &  0.289 &  0.223 &  0.0472 &  1.056 &  0.951 \\  
  65 &  10 29 09.30 &  39 44 26.0 &   1430 53002   389 &  2.460 &  0.0306 &  5.115 &  0.264 &  0.185 &  0.0397 &  1.066 &  0.932 \\  
  66 &  10 33 28.52 &  07 08 01.7 &    999 52636   517 &  2.327 &  0.0168 &  3.464 &  0.216 &  0.175 &  0.0361 &  0.998 &  0.914 \\  
  67 &  10 42 40.51 &  37 55 02.6 &   1998 53433   490 &  2.079 &  0.0393 &  5.497 &  0.216 &  0.168 &  0.0521 &  1.132 &  1.189 \\  
  68 &  10 45 20.41 &  09 23 49.1 &   2147 53491   514 &  1.916 &  0.0472 &  6.212 &  0.208 &  0.158 &  0.0511 &  1.156 &  1.237 \\  
  69 &  10 45 54.47 &  01 04 05.7 &    275 51910   445 &  1.924 &  0.0447 &  6.422 &  0.184 &  0.137 &  0.0424 &  1.122 &  1.100 \\  
  70 &  10 47 23.60 &  30 21 44.2 &   1981 53463   438 &  2.313 &  0.0350 &  5.820 &  0.192 &  0.145 &  0.0415 &  1.068 &  0.983 \\  
  71 &  10 51 08.88 &  13 19 27.9 &   1749 53357   026 &  2.914 &  0.0388 &  5.020 &  0.337 &  0.241 &  0.0543 &  1.164 &  1.002 \\  
  72 &  10 53 28.95 &  43 30 26.1 &   1362 53050   466 &  3.288 &  0.0148 &  3.112 &  0.343 &  0.255 &  0.0417 &  0.993 &  0.834 \\  
  73 &  10 56 39.16 &  67 10 48.9 &    490 51929   401 &  3.263 &  0.0216 &  4.736 &  0.290 &  0.208 &  0.0494 &  0.980 &  0.904 \\  
  74 &  10 59 40.96 &  08 00 56.8 &   1003 52641   327 &  1.883 &  0.0266 &  5.874 &  0.168 &  0.139 &  0.0396 &  0.977 &  1.071 \\  
  75 &  11 09 18.04 &  49 47 53.7 &    964 52646   570 &  2.640 &  0.0422 &  5.593 &  0.267 &  0.193 &  0.0485 &  1.153 &  0.995 \\  
  76 &  11 15 39.03 &  15 26 12.2 &   1752 53379   530 &  2.765 &  0.0378 &  5.300 &  0.303 &  0.240 &  0.0653 &  1.131 &  1.146 \\  
  77 &  11 15 58.16 &  55 48 06.5 &    908 52373   027 &  2.866 &  0.0438 &  5.396 &  0.273 &  0.192 &  0.0456 &  1.184 &  0.925 \\  
  78 &  11 18 48.62 &  30 36 32.2 &   1979 53431   240 &  2.486 &  0.0397 &  5.627 &  0.255 &  0.198 &  0.0524 &  1.127 &  1.072 \\  
  79 &  11 19 59.99 &  34 38 00.3 &   2111 53467   232 &  2.359 &  0.0437 &  6.856 &  0.269 &  0.166 &  0.0375 &  1.089 &  0.925 \\  
  80 &  11 25 46.79 &  47 00 00.3 &   1442 53050   396 &  2.669 &  0.0197 &  4.742 &  0.230 &  0.164 &  0.0433 &  0.952 &  0.934 \\  
  81 &  11 26 40.81 &  35 20 40.6 &   2111 53467   018 &  3.265 &  0.0218 &  4.339 &  0.331 &  0.247 &  0.0566 &  1.008 &  0.966 \\  
  82 &  11 29 14.28 &  36 44 55.9 &   2113 53468   430 &  2.935 &  0.0206 &  5.946 &  0.230 &  0.168 &  0.0430 &  0.905 &  0.890 \\  
  83 &  11 33 10.56 &  30 57 17.6 &   1974 53430   200 &  2.755 &  0.0286 &  5.263 &  0.279 &  0.204 &  0.0547 &  1.033 &  1.037 \\  
  84 &  11 35 30.91 &  11 17 17.8 &   1607 53083   205 &  2.391 &  0.0402 &  5.969 &  0.252 &  0.186 &  0.0418 &  1.109 &  0.970 \\  
  85 &  11 37 06.18 & -03 37 37.1 &    327 52294   042 &  2.216 &  0.0449 &  6.266 &  0.217 &  0.158 &  0.0479 &  1.133 &  1.088 \\  
  86 &  11 37 07.19 &  37 48 33.7 &   2036 53446   598 &  2.565 &  0.0285 &  5.616 &  0.224 &  0.155 &  0.0493 &  1.012 &  1.019 \\  
  87 &  11 38 09.69 &  33 58 05.0 &   2098 53460   503 &  2.206 &  0.0307 &  5.426 &  0.221 &  0.154 &  0.0367 &  1.047 &  0.946 \\  
  88 &  11 40 47.42 &  64 47 10.2 &    597 52059   586 &  2.411 &  0.0478 &  6.410 &  0.215 &  0.143 &  0.0553 &  1.149 &  1.126 \\  
  89 &  11 43 06.52 &  68 07 17.7 &    492 51955   449 &  2.657 &  0.0349 &  5.033 &  0.301 &  0.218 &  0.0532 &  1.121 &  1.042 \\  
  90 &  11 43 33.10 &  53 30 00.6 &   1015 52734   003 &  2.470 &  0.0303 &  5.378 &  0.234 &  0.172 &  0.0416 &  1.046 &  0.952 \\  
  91 &  11 49 04.74 &  15 41 01.8 &   1761 53376   601 &  3.378 &  0.0416 &  5.193 &  0.300 &  0.230 &  0.0583 &  1.178 &  0.963 \\  
  92 &  11 51 49.13 &  31 30 14.5 &   1991 53446   200 &  3.118 &  0.0388 &  4.826 &  0.323 &  0.207 &  0.0488 &  1.180 &  0.918 \\  
  93 &  11 52 30.06 &  07 07 03.9 &   1622 53385   193 &  2.857 &  0.0298 &  4.562 &  0.321 &  0.223 &  0.0500 &  1.098 &  0.970 \\  
  94 &  11 54 31.91 &  08 34 37.5 &   1622 53385   403 &  2.043 &  0.0434 &  5.885 &  0.202 &  0.152 &  0.0453 &  1.144 &  1.104 \\  
  95 &  11 56 05.17 &  14 21 31.5 &   1763 53463   225 &  2.777 &  0.0326 &  4.792 &  0.272 &  0.216 &  0.0639 &  1.113 &  1.128 \\  
  96 &  12 07 25.64 &  62 34 57.9 &    778 52337   498 &  2.284 &  0.0208 &  4.195 &  0.247 &  0.196 &  0.0375 &  1.004 &  0.940 \\  
  97 &  12 11 19.64 &  38 23 30.8 &   2108 53473   077 &  2.280 &  0.0222 &  4.839 &  0.210 &  0.158 &  0.0404 &  0.981 &  0.976 \\  
  98 &  12 13 49.66 &  61 25 29.0 &    779 52342   218 &  2.646 &  0.0227 &  4.784 &  0.272 &  0.199 &  0.0512 &  0.991 &  1.023 \\  
  99 &  12 20 40.85 &  33 31 28.0 &   1999 53503   119 &  2.684 &  0.0184 &  3.764 &  0.292 &  0.216 &  0.0429 &  1.000 &  0.927 \\  
 100 &  12 24 41.76 &  14 58 04.6 &   1767 53436   514 &  2.282 &  0.0399 &  5.865 &  0.222 &  0.165 &  0.0587 &  1.113 &  1.209 \\  
 101 &  12 25 57.50 &  39 21 56.9 &   1986 53475   136 &  2.325 &  0.0497 &  6.578 &  0.231 &  0.165 &  0.0684 &  1.154 &  1.323 \\  
 102 &  12 27 20.15 &  51 39 43.3 &    884 52374   216 &  2.521 &  0.0357 &  5.708 &  0.256 &  0.191 &  0.0417 &  1.082 &  0.943 \\  
 103 &  12 28 08.06 &  07 54 43.4 &   1627 53473   426 &  2.130 &  0.0544 &  7.139 &  0.188 &  0.150 &  0.0529 &  1.158 &  1.182 \\  
 104 &  12 48 58.09 &  47 01 37.4 &   1455 53089   076 &  2.617 &  0.0454 &  5.949 &  0.258 &  0.188 &  0.0503 &  1.159 &  1.019 \\  
 105 &  12 49 54.84 &  06 06 10.3 &    847 52426   522 &  2.615 &  0.0404 &  5.288 &  0.267 &  0.190 &  0.0383 &  1.159 &  0.890 \\  
 106 &  12 52 14.32 &  00 51 58.4 &    292 51609   566 &  2.537 &  0.0419 &  6.432 &  0.214 &  0.167 &  0.0535 &  1.097 &  1.072 \\  
 107 &  12 53 21.75 &  50 24 17.8 &   1279 52736   546 &  2.044 &  0.0492 &  6.817 &  0.188 &  0.146 &  0.0468 &  1.136 &  1.125 \\  
 108 &  12 53 28.60 &  58 40 21.9 &    957 52398   210 &  2.060 &  0.0536 &  7.109 &  0.189 &  0.147 &  0.0494 &  1.153 &  1.157 \\  
 109 &  12 57 34.42 &  15 22 29.2 &   1771 53498   429 &  2.839 &  0.0274 &  4.183 &  0.345 &  0.245 &  0.0662 &  1.098 &  1.137 \\  
 110 &  13 01 19.26 &  12 39 59.4 &   1695 53473   627 &  2.280 &  0.0231 &  4.440 &  0.212 &  0.175 &  0.0424 &  1.019 &  1.001 \\  
 111 &  13 02 04.39 &  42 48 12.3 &   1458 53119   543 &  3.359 &  0.0244 &  3.782 &  0.357 &  0.259 &  0.0504 &  1.093 &  0.900 \\  
 112 &  13 07 28.69 &  54 26 49.6 &   1039 52707   119 &  2.701 &  0.0342 &  6.376 &  0.196 &  0.160 &  0.0518 &  1.030 &  1.018 \\  
 113 &  13 12 35.15 &  12 57 43.3 &   1697 53142   415 &  2.381 &  0.0477 &  6.731 &  0.206 &  0.156 &  0.0489 &  1.129 &  1.057 \\  
 114 &  13 16 44.77 &  10 57 32.9 &   1697 53142   048 &  1.839 &  0.0536 &  7.344 &  0.178 &  0.134 &  0.0456 &  1.140 &  1.181 \\  
 115 &  13 20 32.02 &  40 59 01.5 &   1462 53112   184 &  2.591 &  0.0278 &  4.982 &  0.195 &  0.165 &  0.0547 &  1.042 &  1.073 \\  
 116 &  13 21 14.47 &  46 05 04.7 &   1461 53062   068 &  2.487 &  0.0299 &  5.112 &  0.196 &  0.148 &  0.0431 &  1.059 &  0.965 \\  
 117 &  13 21 51.93 &  47 08 35.9 &   1461 53062   582 &  2.262 &  0.0603 &  6.996 &  0.224 &  0.171 &  0.0540 &  1.211 &  1.154 \\  
 118 &  13 24 55.03 &  57 45 10.6 &    959 52411   137 &  1.699 &  0.0644 &  7.740 &  0.139 &  0.109 &  0.0466 &  1.194 &  1.261 \\  
 119 &  13 26 38.45 &  42 14 53.6 &   1462 53112   608 &  3.031 &  0.0163 &  3.318 &  0.330 &  0.240 &  0.0425 &  1.002 &  0.872 \\  
 120 &  13 27 38.26 &  32 09 51.4 &   2110 53467   409 &  2.563 &  0.0165 &  3.312 &  0.244 &  0.190 &  0.0321 &  1.006 &  0.829 \\  
 121 &  13 28 44.05 &  43 55 50.5 &   1376 53089   637 &  2.209 &  0.0336 &  5.466 &  0.193 &  0.150 &  0.0433 &  1.075 &  1.030 \\  
 122 &  13 29 23.46 & -03 15 02.2 &    911 52426   253 &  2.390 &  0.0266 &  4.708 &  0.252 &  0.183 &  0.0441 &  1.047 &  0.997 \\  
 123 &  13 34 44.88 &  59 44 34.2 &    960 52425   368 &  2.899 &  0.0354 &  4.383 &  0.333 &  0.246 &  0.0490 &  1.182 &  0.953 \\  
 124 &  13 34 44.88 &  59 44 34.2 &    960 52466   364 &  3.284 &  0.0311 &  4.256 &  0.324 &  0.230 &  0.0529 &  1.140 &  0.931 \\  
 125 &  13 37 02.29 &  05 29 11.3 &    853 52374   540 &  2.488 &  0.0305 &  5.305 &  0.292 &  0.209 &  0.0579 &  1.053 &  1.136 \\  
\hline 
\end{tabular}\\
\end{center}
\end{table*}

%+++++++++++++++++++++++++++++++++ Table 1    Lines in the SDSS sample
\setcounter{table}{0}
\begin{table*}
%\rotate
%\tabletypesize{\tiny}
%\tabletypesize{\scriptsize}
%\tabletypesize{\small}
%\tabletypesize{\normalsize}
\caption[]{
Continued.
}
%\begin{flushleft}
\begin{center}
\begin{tabular}{rrrrrrcccccc} \hline \hline
No$^a$                                                   & 
RA$^b$                                                   & 
DEC$^c$                                                  & 
Spectrum$^d$                                             & 
[O\,{\sc ii}]                                            & 
[O\,{\sc iii}]                                           & 
[O\,{\sc iii}]                                           & 
[S\,{\sc ii}]                                            & 
[S\,{\sc ii}]                                            & 
[O\,{\sc ii}]                                            & 
t$_{\rm 3,O}$                                            & 
t$_{\rm 2,O}$                                            \\
                                                         & 
                                                         & 
                                                         & 
number                                                   & 
$\lambda$3727                                            & 
$\lambda$4363                                            & 
$\lambda$4959+$\lambda$5007                              & 
$\lambda$6717                                            & 
$\lambda$6731                                            & 
$\lambda$7325                                            & 
                                                         & 
                                                         \\   \hline
 126 &  13 37 57.46 &  12 09 41.6 &   1700 53502   461 &  2.447 &  0.0470 &  6.096 &  0.261 &  0.189 &  0.0564 &  1.162 &  1.129 \\  
 127 &  13 40 49.08 &  38 24 44.2 &   2005 53472   106 &  2.293 &  0.0121 &  3.324 &  0.212 &  0.167 &  0.0403 &  0.919 &  0.972 \\  
 128 &  13 41 06.77 &  60 53 47.5 &    785 52339   079 &  3.509 &  0.0199 &  4.104 &  0.298 &  0.237 &  0.0555 &  0.997 &  0.923 \\  
 129 &  13 42 52.05 &  05 14 15.2 &    854 52373   373 &  3.021 &  0.0189 &  4.861 &  0.277 &  0.215 &  0.0522 &  0.935 &  0.964 \\  
 130 &  13 44 27.36 &  56 01 29.7 &   1321 52764   624 &  2.484 &  0.0449 &  6.570 &  0.217 &  0.174 &  0.0567 &  1.115 &  1.123 \\  
 131 &  13 44 27.36 &  56 01 29.7 &   1322 52791   470 &  2.679 &  0.0420 &  6.544 &  0.217 &  0.178 &  0.0618 &  1.091 &  1.130 \\  
 132 &  13 48 06.98 &  26 24 19.6 &   2115 53535   414 &  2.432 &  0.0665 &  7.309 &  0.191 &  0.144 &  0.0483 &  1.235 &  1.037 \\  
 133 &  13 50 43.51 &  24 34 24.6 &   2115 53535   215 &  3.488 &  0.0142 &  3.551 &  0.369 &  0.272 &  0.0356 &  0.942 &  0.762 \\  
 134 &  13 52 20.71 &  38 48 49.7 &   2014 53460   525 &  3.024 &  0.0344 &  4.507 &  0.321 &  0.225 &  0.0527 &  1.158 &  0.968 \\  
 135 &  13 52 48.26 &  11 14 10.6 &   1702 53144   234 &  2.910 &  0.0254 &  5.169 &  0.264 &  0.189 &  0.0457 &  1.001 &  0.920 \\  
 136 &  13 56 24.45 &  57 45 47.1 &   1158 52668   120 &  3.900 &  0.0321 &  4.457 &  0.324 &  0.256 &  0.0633 &  1.135 &  0.934 \\  
 137 &  14 09 34.94 &  06 10 23.6 &   1824 53491   063 &  2.133 &  0.0452 &  7.085 &  0.187 &  0.137 &  0.0530 &  1.089 &  1.182 \\  
 138 &  14 10 07.10 &  45 08 17.5 &   1467 53115   579 &  3.254 &  0.0282 &  4.160 &  0.356 &  0.267 &  0.0466 &  1.112 &  0.880 \\  
 139 &  14 10 22.26 &  44 14 55.8 &   1467 53115   032 &  2.848 &  0.0357 &  4.979 &  0.308 &  0.208 &  0.0505 &  1.134 &  0.977 \\  
 140 &  14 16 43.53 &  09 00 50.6 &   1811 53533   347 &  2.756 &  0.0329 &  5.187 &  0.308 &  0.233 &  0.0681 &  1.087 &  1.178 \\  
 141 &  14 27 09.15 &  12 58 45.5 &   1708 53503   512 &  2.919 &  0.0271 &  4.573 &  0.273 &  0.206 &  0.0500 &  1.063 &  0.960 \\  
 142 &  14 31 04.65 &  23 00 48.1 &   2136 53494   265 &  2.227 &  0.0286 &  5.942 &  0.238 &  0.185 &  0.0699 &  0.996 &  1.386 \\  
 143 &  14 40 57.31 &  12 20 12.9 &   1710 53504   632 &  2.563 &  0.0302 &  4.646 &  0.326 &  0.228 &  0.0581 &  1.096 &  1.118 \\  
 144 &  14 48 25.52 &  63 10 10.6 &    609 52339   454 &  2.481 &  0.0379 &  5.709 &  0.259 &  0.190 &  0.0498 &  1.104 &  1.043 \\  
 145 &  14 48 39.66 &  54 40 05.0 &   1163 52669   465 &  2.423 &  0.0316 &  5.633 &  0.252 &  0.198 &  0.0616 &  1.044 &  1.200 \\  
 146 &  15 09 09.03 &  45 43 08.8 &   1050 52721   274 &  2.130 &  0.0468 &  6.701 &  0.162 &  0.129 &  0.0386 &  1.123 &  0.988 \\  
 147 &  15 10 28.60 &  34 16 18.4 &   1385 53108   559 &  2.882 &  0.0213 &  4.718 &  0.338 &  0.240 &  0.0495 &  0.976 &  0.961 \\  
 148 &  15 12 12.85 &  47 16 30.7 &   1050 52721   402 &  2.271 &  0.0316 &  5.197 &  0.224 &  0.172 &  0.0446 &  1.071 &  1.031 \\  
 149 &  15 16 34.76 &  30 06 53.6 &   1650 53174   305 &  2.573 &  0.0337 &  5.435 &  0.234 &  0.181 &  0.0490 &  1.079 &  1.014 \\  
 150 &  15 28 17.18 &  39 56 50.4 &   1293 52765   580 &  2.168 &  0.0500 &  6.913 &  0.196 &  0.153 &  0.0553 &  1.137 &  1.202 \\  
 151 &  15 28 21.97 &  36 24 09.4 &   1401 53144   397 &  1.993 &  0.0458 &  7.154 &  0.176 &  0.133 &  0.0463 &  1.090 &  1.135 \\  
 152 &  15 30 41.26 &  31 01 06.5 &   1388 53119   189 &  2.076 &  0.0443 &  6.578 &  0.200 &  0.154 &  0.0482 &  1.109 &  1.134 \\  
 153 &  15 36 56.44 &  31 22 48.0 &   1388 53119   039 &  2.914 &  0.0347 &  5.404 &  0.250 &  0.174 &  0.0439 &  1.092 &  0.902 \\  
 154 &  15 37 53.31 &  58 41 37.7 &    615 52347   590 &  2.966 &  0.0448 &  5.987 &  0.268 &  0.197 &  0.0581 &  1.150 &  1.029 \\  
 155 &  15 46 13.86 &  32 56 32.3 &   1580 53145   418 &  2.241 &  0.0353 &  5.526 &  0.242 &  0.173 &  0.0483 &  1.089 &  1.086 \\  
 156 &  15 46 13.86 &  32 56 32.3 &   1403 53227   302 &  2.520 &  0.0340 &  5.169 &  0.228 &  0.184 &  0.0579 &  1.100 &  1.127 \\  
 157 &  15 58 22.20 &  36 11 43.7 &   1682 53173   277 &  2.368 &  0.0370 &  6.074 &  0.226 &  0.178 &  0.0532 &  1.073 &  1.112 \\  
 158 &  15 58 42.48 &  17 21 37.5 &   2196 53534   331 &  2.776 &  0.0354 &  5.894 &  0.224 &  0.173 &  0.0492 &  1.068 &  0.977 \\  
 159 &  16 01 35.95 &  31 13 53.7 &   1405 52826   395 &  3.050 &  0.0351 &  5.485 &  0.197 &  0.150 &  0.0426 &  1.090 &  0.870 \\  
 160 &  16 02 03.58 &  29 26 14.2 &   1578 53496   438 &  2.919 &  0.0298 &  5.189 &  0.291 &  0.215 &  0.0533 &  1.052 &  0.992 \\  
 161 &  16 15 53.92 &  27 11 43.5 &   1576 53496   378 &  2.335 &  0.0376 &  5.145 &  0.262 &  0.188 &  0.0449 &  1.140 &  1.019 \\  
 162 &  16 18 32.65 &  27 43 52.4 &   1576 53496   446 &  1.886 &  0.0265 &  6.272 &  0.181 &  0.138 &  0.0395 &  0.957 &  1.068 \\  
 163 &  16 24 10.10 & -00 22 02.5 &    364 52000   187 &  1.468 &  0.0582 &  7.933 &  0.127 &  0.092 &  0.0359 &  1.142 &  1.171 \\  
 164 &  16 24 21.38 &  27 04 08.7 &   1408 52822   221 &  2.441 &  0.0307 &  5.352 &  0.243 &  0.182 &  0.0442 &  1.051 &  0.987 \\  
 165 &  16 29 34.81 &  30 33 29.8 &   1685 53463   234 &  2.719 &  0.0213 &  4.746 &  0.297 &  0.208 &  0.0428 &  0.975 &  0.921 \\  
 166 &  16 35 27.94 &  22 25 18.8 &   1571 53174   155 &  2.455 &  0.0273 &  5.161 &  0.260 &  0.191 &  0.0428 &  1.025 &  0.968 \\  
 167 &  16 47 25.17 &  30 27 29.2 &   1342 52793   537 &  2.745 &  0.0212 &  4.070 &  0.263 &  0.204 &  0.0460 &  1.020 &  0.949 \\  
 168 &  16 59 43.89 &  24 55 50.6 &   1693 53446   198 &  2.354 &  0.0229 &  4.941 &  0.324 &  0.244 &  0.0448 &  0.985 &  1.014 \\  
 169 &  17 03 05.08 &  25 31 47.6 &   1693 53446   541 &  3.259 &  0.0254 &  3.682 &  0.318 &  0.240 &  0.0412 &  1.119 &  0.833 \\  
 170 &  17 24 37.52 &  56 28 37.7 &    367 51997   561 &  3.289 &  0.0281 &  5.375 &  0.291 &  0.208 &  0.0611 &  1.021 &  1.001 \\  
 171 &  17 24 37.52 &  56 28 37.7 &    357 51813   568 &  3.575 &  0.0392 &  5.573 &  0.302 &  0.221 &  0.0576 &  1.126 &  0.931 \\  
 172 &  21 04 15.71 & -06 05 17.3 &    637 52174   526 &  2.482 &  0.0272 &  5.242 &  0.246 &  0.185 &  0.0458 &  1.019 &  0.997 \\  
 173 &  21 15 27.07 & -07 59 51.4 &    639 52146   242 &  1.757 &  0.0357 &  6.265 &  0.183 &  0.134 &  0.0348 &  1.050 &  1.036 \\  
 174 &  21 39 56.50 &  00 19 21.7 &   1108 53227   397 &  1.874 &  0.0531 &  6.857 &  0.232 &  0.178 &  0.0527 &  1.164 &  1.282 \\  
 175 &  21 46 42.28 &  00 00 08.9 &   1030 52914   077 &  1.924 &  0.0506 &  7.041 &  0.175 &  0.139 &  0.0368 &  1.134 &  1.016 \\  
 176 &  22 15 89.25 &  01 09 39.0 &   1034 52525   551 &  2.651 &  0.0487 &  5.654 &  0.256 &  0.196 &  0.0488 &  1.210 &  0.996 \\  
 177 &  22 15 89.25 &  01 09 39.0 &   1034 52813   521 &  2.369 &  0.0435 &  5.544 &  0.272 &  0.192 &  0.0420 &  1.170 &  0.977 \\  
 178 &  22 15 89.25 &  01 09 39.0 &   1476 52964   530 &  2.576 &  0.0498 &  5.631 &  0.267 &  0.197 &  0.0423 &  1.222 &  0.940 \\  
 179 &  22 51 40.31 &  12 27 13.3 &    741 52261   279 &  2.490 &  0.0395 &  5.582 &  0.238 &  0.176 &  0.0422 &  1.128 &  0.955 \\  
 180 &  23 29 36.55 & -01 10 56.9 &    384 51821   281 &  2.057 &  0.0650 &  7.554 &  0.196 &  0.138 &  0.0433 &  1.210 &  1.071 \\  
 181 &  23 29 36.55 & -01 10 56.9 &    681 52199   201 &  1.806 &  0.0670 &  7.680 &  0.191 &  0.146 &  0.0422 &  1.216 &  1.139 \\  
\hline 
\end{tabular}\\
\end{center}
\begin{flushleft}
$^a$ The objects are listed in order of right ascension. \\
$^b$ Units of right ascension are hours, minutes, and seconds. \\
$^c$ Units of declination are degrees, arcminutes, and arcseconds. \\
$^d$ The spectrum number is composed of the SDSS plate number, 
the modified Julian date of the observation, 
and the number of the fiber on the plate.
\end{flushleft}
\end{table*}

\end{document}